\documentclass[twocolumn]{aastex631}
\usepackage{soul}

\usepackage{xcolor}

\definecolor{BrightMaroon}{RGB}{194, 24, 91}

\definecolor{BrightPurple}{RGB}{147, 0, 255}  

\definecolor{Sepia}{RGB}{112, 66, 20}  

\definecolor{BlueGreen}{RGB}{13, 152, 186}  

\usepackage{newtxtext,newtxmath}
\usepackage{amsmath}
\usepackage{latexsym}
\usepackage{amssymb}	
\usepackage{gensymb}
\usepackage{flafter}
\usepackage{appendix}
\usepackage{natbib}
\graphicspath{{./}{figures/}}

\usepackage{subfigure}
\usepackage[figuresright]{rotating}
\usepackage{makecell}
\usepackage{booktabs}
\usepackage{multirow}
\usepackage{threeparttable}
\usepackage{color}

\begin{document}

\title{Spectral evolution of GX 17+2 using AstroSat and  NICER observations}
\shorttitle{GX 17+2: Spectral analysis}
\shortauthors{Bhattacherjee et al.}

\author[0000-0002-1418-1856]{Sree Bhattacherjee}
\affiliation{Department of Applied Sciences, Tezpur University, Napaam, Assam 784028, India}

\author[0000-0002-7609-2779]{Ranjeev Misra}
\affiliation{Inter University Center for Astronomy and Astrophysics, Ganeshkhind, Pune 411007, India}

\author[0000-0003-4402-0970]{Biplob Sarkar}
\affiliation{Department of Applied Sciences, Tezpur University, Napaam, Assam 784028, India}

\correspondingauthor{}
\email{E-mail: biplobs@tezu.ernet.in}

\author[0000-0002-6449-9643]{V. Jithesh}
\affiliation{Department of Physics and Electronics, Christ University, Hosur Main Road, Bengaluru - 560029, India}

\author[0000-0002-2329-5863]{Jayashree Roy }
\affiliation{Inter University Center for Astronomy and Astrophysics, Ganeshkhind, Pune 411007, India}

\author{Yashpal Bhulla}
\affiliation{Pacific Academy of Higher Education and Research University, Udaipur 313003, India}

\begin{abstract}
We study the spectral evolution of the Z-track source GX 17+2 using AstroSat and NICER observations taken between 2016 and 2020. The AstroSat observations cover the period when the source is in the normal branch (NB) and the flaring branch (FB), while for the NICER ones the variability can be associated with the FB branch. The source spectra at different regions of the branches are well described by accretion disk emission, blackbody surface emission and a thermal Comptonization component. In the NB, the  total bolometric unabsorbed flux  remains constant and the variation is due to changes in the Comptonization, disk fluxes. In particular, the inferred luminosity ($L_{\rm T}$) and accretion rate ($\dot M$) remain constant, while there is significant variation in the inner disk radii and fraction of disk photons entering the corona, indicating changes in the geometry of the system. On the other hand, in the FB, there is significant variation in luminosity from $\sim 4.0$ to $\sim 7.0 \times 10^{38}$ ergs s$^{-1}$. Despite this significant variation in luminosity and in the inner disk radii, the accretion efficiency defined as $\eta = L_{\rm T}/{\dot M} c^2$, remains nearly constant at $\sim 0.20$ throughout the evolution of the source, as expected for a neutron star system.

\end{abstract}


\section{Introduction} \label{sec:intro}

Neutron star low-mass X-ray binary (NS-LMXB) can be classified into Z-type and atoll-type sources, based on their correlated spectral and timing properties. Z-track sources are named so since they trace a `Z'-like shape in their hardness-intensity diagram (HID) and color-color diagram (CCD). Whereas  atoll-type sources trace out a `C'-shaped HID and CCD \citep{1989Hasinger}.  Z-sources have higher luminosities compared to atoll sources. Essentially, their luminosities are almost equivalent to the Eddington limit; hence, this class of NS-LMXB counts as one of the most luminous and persistent X-ray sources that accrete mass near the Eddington rate (e.g.,  \citet{1989Lamb}). In Z-sources, the three branches from top to bottom are called the horizontal branch (HB), normal branch (NB), and flaring branch (FB), which corresponds to three distinct spectral states of the source. The Z-sources trace smoothly from HB to NB via the hard apex (HA) and from NB to FB via the soft apex (SA). Based on the Z-track pattern traced in the HID, magnetic field strength and luminosity, these sources are further categorized into Sco-like (Sco X-1, GX 17+2, and GX 349+2) and Cyg-like (Cyg X-2, GX 340+0, and GX 5-1) sources \citep{1989Hasinger,1994Kuulkers, 1997Kuulkers}. Cyg-like sources have more dominant HB than FB, while Sco-like sources have a well-pronounced FB but a small or no HB \citep{2012Church}. 

Z-sources have a short timescale of variability from hours to days \citep{1989Hasinger, 2002Piraino}, keeping them consistently near the threshold of a state change. This makes them excellent subjects to study the correlations between state transitions and X-ray properties in X-ray binary systems. Despite extensive studies on Z-track LMXBs, the factors influencing a source's spectral state and its position on CCD or HID are still under debate. It is yet unknown what mechanisms lead the source to transition between states and follow its Z-track. Traditionally, changes in the mass accretion rate ($\dot{M}$) have been considered responsible for such transitions, as suggested by \citet{1986Priedhorsky}. It is commonly believed that $\dot{M}$ increase from the HB --> NB --> FB \citep{1990Hasinger, 1990Vrtilek}. However, some recent findings indicate that the variability of Z sources is more complex than depending on a single factor \citep{2009Lin}. \citet{2002Homan, 2012Lin} suggested that the $\dot{M}$ might be rather consistent along the Z track. According to \citet{2012Lin}, the movement along the Z track may be explained by instabilities in the accretion disk. While, \citet{2006Church} claimed that the FB is caused by variation in the thermonuclear burning rate and  the $\dot{M}$ increases from the NB to the HB. Hence, such perplexity highlights the requirement for rigorous comprehension of the physical parameters among the different spectral states of Z sources. 
\begin{figure}
	\centering
	\includegraphics[scale=0.38]{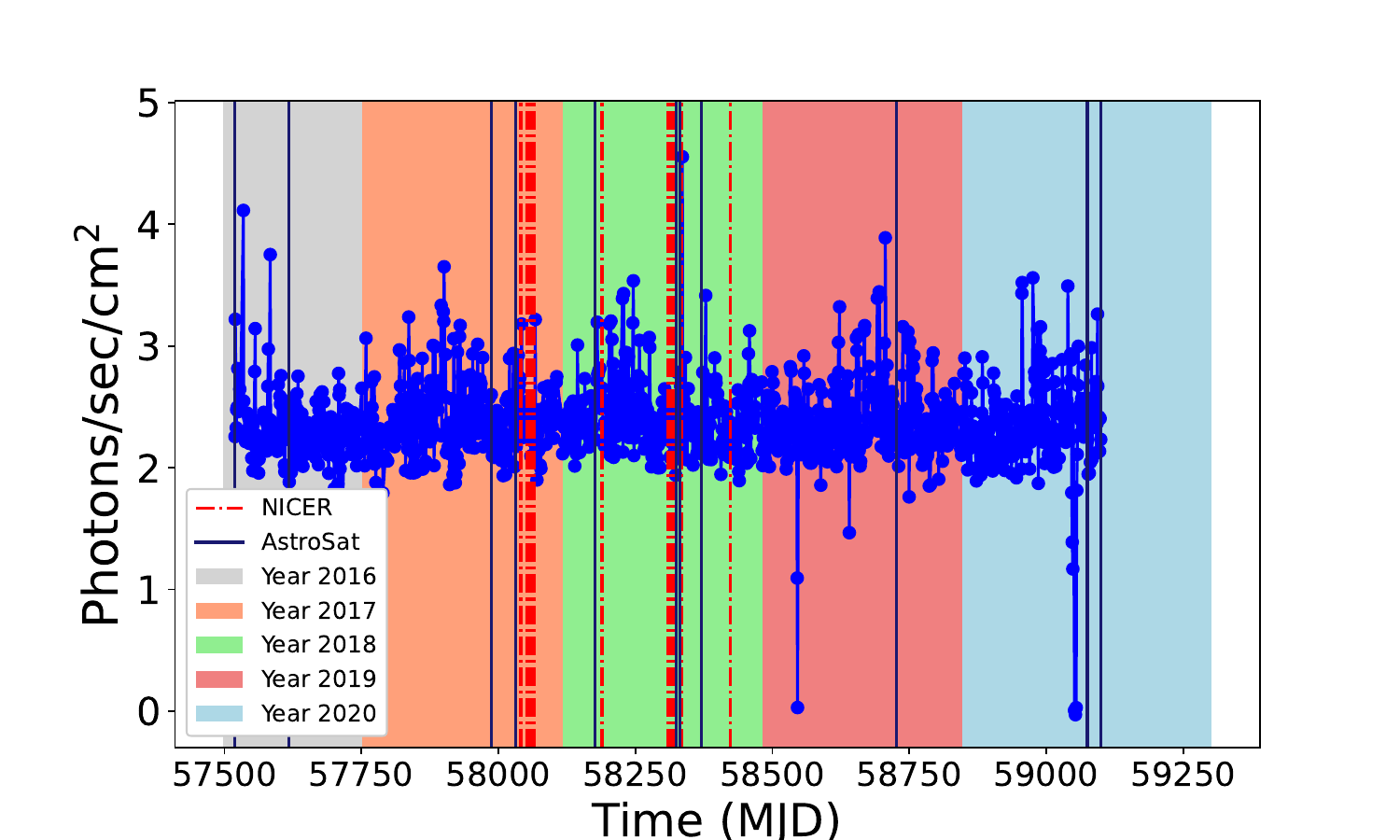}
    \caption{The one-day binned MAXI (2-20 keV) longterm lightcurve of GX 17+2 spanning from the year 2016 to 2020. The solid line represents the AstroSat observations, and the dashed line marks the NICER observations for the source. }
    \label{fig:maxi}
\end{figure}

The X-ray spectra of Z-sources are generally soft in all three branches and hence are predominated by soft/thermal constituents, with the majority of the flux emitted below 20 keV. The soft state spectra are typically represented by models that incorporate both a thermal component and a Comptonized component. Depending on the selection of these components, there are two traditional models known as the Eastern model \citep{1984Mitsuda, 1986Makishima, 1989Mitsuda}, described by multicolor disk blackbody for soft component emission with a weak Comptonized blackbody component, and the Western model \citep{1988White}, where single-color blackbody is responsible for the soft component emission associated with Comptonized emission from the disk.

The bright NS-LMXB, GX 17+2, was discovered in the Sagittarius constellation by an Aerobee 150 sounding rocket through several experiments during 1965 \citep{1967Gursky, 1967Friedman, 1968Bradt,1971Tananbaum}. It is situated at a distance of 13 kpc \citep{2008Galloway} and has a spin frequency of 293.2 Hz \citep{1997Wijnands}. It is a relatively low inclination system with $\theta$$\sim$ 30$^{\circ}$-40$^{\circ}$ \citep{2010Cackett, 2017Ludlam}. Several X-ray spectral studies have been carried out for GX 17+2 \citep{2000DiSalvo, 2002Homan,2005Farinelli, 2010Cackett,2012Lin}. \citet{2000DiSalvo} carried out a detailed spectral analysis of GX 17+2 for the first time. Using data from BeppoSAX (0.1–200 keV), authors studied the X-ray spectrum and observed a prominent hard X-ray tail in the HB compared to other branches. This finding indicates that the hard tail may result from the Comptonization of seed photons in a relativistic jet. Furthermore, it is proposed that the X-ray emission is dependent on the rate of accretion. Later on, \citet{2005Farinelli, 2007Migliari} also reported the hard X-ray tail signature in HB. \citet{2002Homan} performed a comprehensive spectral and timing study using $\sim$600 ks  Rossi X-Ray Timing Explorer (RXTE) data, demonstrating the co-relation of quasi-periodic oscillations (QPOs) with the spectral states. They also compared the source to black hole LMXBs and speculated that $\dot{M}$ might not be the underlying cause for spectral state transitions. Recently, using AstroSat observations, \citet{2020Malu} observed  normal branch oscillations (NBOs) at frequency $\sim$6.7--7.8 Hz. Similarly, \citet{2020Agrawal} also detected NBOs at a centroid frequency of 7.42$\pm$0.23 Hz. Along with timing variability, the authors also performed an extensive spectral analysis using various models. They observed that while the strength of the power-law component depends on the model, it decreases from HB to NB and again starts increasing from NB to FB. While \citet{2021Sriram} detected horizontal branch oscillations (HBOs) at $\sim$  25 Hz and $\sim$  33 Hz, they also included spectral analysis for the segments associated with significant lags. However, no notable variations in any spectral parameters were found; instead, a few sections showed slight variations in flux. Additionally, the inner disk radius was found to be near the final stable orbit along the HB-NB/FB, indicating that the disk is not truncated.

In order to comprehend the behavior of the source along the branches, \citet{2012Lin} performed the S$_z$ resolved spectroscopy of GX 17+2 using the RXTE data. The authors predicted that the evolution of each branch of the Z-track could be driven by three different factors with a constant $\dot{M}$. The three factors being: increase in Comptonization with respect to the upper vertex, resulting in the formation of HB. From the lower vertex, the inner disk radius rapidly shrinks, resulting in the FB formation, while in NB, changes in the boundary layer area was observed indicating a change in accretion from slim disk at the upper vertex to a thin disk in the upper vertex. 

In this work, we study the spectral evolution of the Sco-like GX 17+2 source using the publicly available AstroSat and Neutron Star Interior Composition Explorer (NICER) observations from the year 2016 to 2020, which provide us the opportunity to probe the source in broad-band spectra from soft to hard X-ray energy range. We use the flux-resolved spectroscopy method to understand the behavior of the source with respect to the different intensity states. Moreover, such a flux correlation study has been done using the AstroSat observations on some of the other Z-sources like GX 340+0 \citep{2023Bhargava, 2024Chattopadhyay}, GX 5-1 \citep{2019Bhulla}, and GX 349+2 \citep{2023Kashyap}, which we compare with our findings. We use the AstroSat/Large Area X-ray Proportional Counter (LAXPC) and Soft X-ray Telescope (SXT), which provide broad spectral coverage from low to high energies, while NICER offers high sensitivity, low background noise, and good energy resolution in the soft X-ray range. Together, these instruments facilitate to a comprehensive comparison of spectral results. This paper discusses the observation and data reduction methods in Section~\ref{sec:obs}. Section~\ref{sec:sa} describes the flux-resolved spectroscopy of GX 17+2 along the Z-track using AstroSat and NICER observations. Section~\ref{sec:r} deals with the explanation of the obtained results. Finally, we discuss and conclude in  Section~\ref{sec:d}.

\section{Observation and Data Reduction} \label{sec:obs}
We analyzed the AstroSat and NICER publicly available archival data of GX 17+2 from their initial observations till the year 2020. Figure~\ref{fig:maxi} marks the observations over the long term MAXI (2-20 keV) lightcurve showing the persistent behavior of the source.

\begin{table}
	\centering
\caption{\label{tab:1}{Log table for AstroSat and NICER observations of GX 17+2.  }}
\scalebox{0.749}{
\begin{tabular}{cccc}
\hline\hline
 Obs Name & Obs ID & Date (YYYY-MM-DD) & Expo. (ks)\\ 
 \hline
\hline
A1   & G05\textunderscore 112T01\textunderscore9000000452	& 2016-05-11  & $\sim$ 99   \\		
A2   & A03\textunderscore 072T01\textunderscore9000001484 	& 2017-08-22  & $\sim$ 40  \\
A3   & G08\textunderscore 037T01\textunderscore9000001588  & 2017-10-06  & $\sim$ 18\\
A4   & G08\textunderscore 037T01\textunderscore9000001928	 & 2018-02-28  & $\sim$ 18   \\
A5   & G08\textunderscore 037T01\textunderscore9000002256   & 2018-07-26  & $\sim$ 20  \\
A6   & G08\textunderscore 037T01\textunderscore9000002264   & 2018-08-01  & $\sim$ 20  \\
A7$^{\rm(a)}$   & T02\textunderscore 087T01\textunderscore9000002352 	  & 2018-09-10  & $\sim$ 50\\
A8   & A05\textunderscore 062T02\textunderscore9000003138    & 2019-09-01  & $\sim$ 20   \\
A9$^{\rm(b)}$ & A09\textunderscore 044T02\textunderscore9000003814   & 2020-08-14  & $\sim$ 7  \\
A10$^{\rm(c)}$ & A09\textunderscore 044T02\textunderscore9000003852    & 2020-09-08  & $\sim$ 16  \\
N1   	 & 1050410101	& 2017-10-16	  	& $\sim$ 560   \\		
N2  	 & 1050410102	& 2017-10-26 	 	& $\sim$ 942  \\
N3   	 & 1050410103 	& 2017-10-27  	 	& $\sim$ 1353  \\
N4	 & 1050410104	& 2017-10-28 		& $\sim$ 1307\\
N5	 & 1050410105	& 2017-10-29	 	 & $\sim$ 1314  \\
N6	 &1050410106 	& 2017-10-31		  & $\sim$ 413 \\
N7	 & 1050410107 	 & 2017-11-02		& $\sim$ 2962\\
N8	 & 1050410108 	 & 2017-11-03 	 	& $\sim$ 135\\
N9	 & 1050410109       & 2017-11-04  		& $\sim$ 1536  \\
N10 	 & 1050410110 	 & 2017-11-05		& $\sim$ 3939 \\
N11 	 & 1050410111    	 & 2017-11-06 		& $\sim$ 3163 \\
N12	 & 1050410112  	& 2017-11-07	  	& $\sim$ 3208   \\		
N13  	 & 1050410113   	& 2017-11-08		& $\sim$ 1214 \\
N14*   	 & 1050410114    	& 2018-03-11		  & $\sim$ 87  \\
N15*  	 & 1050410115    	& 2018-03-11 		 & $\sim$ 915\\
N16   	 & 1050410116    	& 2018-03-13 		& $\sim$ 1136   \\
N17   	 & 1050410117   	&2018-03-14 		 & $\sim$ 2786  \\
N18  	 & 1050410118    	& 2018-03-15 		 & $\sim$ 2607  \\
N19       & 1050410119    	& 2018-03-16 		& $\sim$ 1243  \\
N20       & 1050410120    	& 2018-07-11 		& $\sim$ 729 \\
N21       & 1050410121    	&2018-07-12 	 	& $\sim$ 3621  \\
N22       & 1050410122    	& 2018-07-13		 & $\sim$ 238  \\
N23       & 1050410123    	& 2018-07-14		 & $\sim$ 1038  \\
N24       & 1050410124  	& 2018-07-15		  & $\sim$ 933\\
N25       & 1050410125     	& 2018-07-16		  & $\sim$ 867   \\
N26       & 1050410126     	& 2018-07-18		  & $\sim$ 138  \\
N27       & 1050410127 	& 2018-07-20		  & $\sim$ 743\\
N28*       & 1050410128     	 & 2018-07-21 		 & $\sim$ 395  \\
N29*       & 1050410129       	& 2018-07-21 		 & $\sim$ 2283 \\
N30       & 1050410130       	& 2018-07-23	 	& $\sim$ 181  \\
N31       & 1050410131       	& 2018-08-05  		& $\sim$ 637 \\
N32$^{\rm(a)}$     & 1050410132         & 2018-09-10  		& $\sim$ 21456 \\
N33       & 1050410133       	&2018-11-02 		& $\sim$ 3849  \\
N34*,$^{\rm(b)}$   & 3050410101        & 2020-08-14 	 	& $\sim$ 1882  \\
N35*,$^{\rm(b)}$     & 3050410102       	& 2020-08-14   		& $\sim$ 6109  \\
N36*,$^{\rm(c)}$    & 3050410103       & 2020-09-08 	 	& $\sim$ 2681  \\
N37*,$^{\rm(c)}$      & 3050410104       	& 2020-09-08 	 	& $\sim$ 8940 \\

\hline
\multicolumn{4}{p{10cm}}{Note: `A' \& `N' represents the AstroSat and NICER observations. *two observations in a particular day, which we merged to a single day observation. **(a),(b),(c): observations having simultaneous data with AstroSat.}
\end{tabular}
}
\end{table}

\begin{figure*}
\includegraphics[scale=0.59]{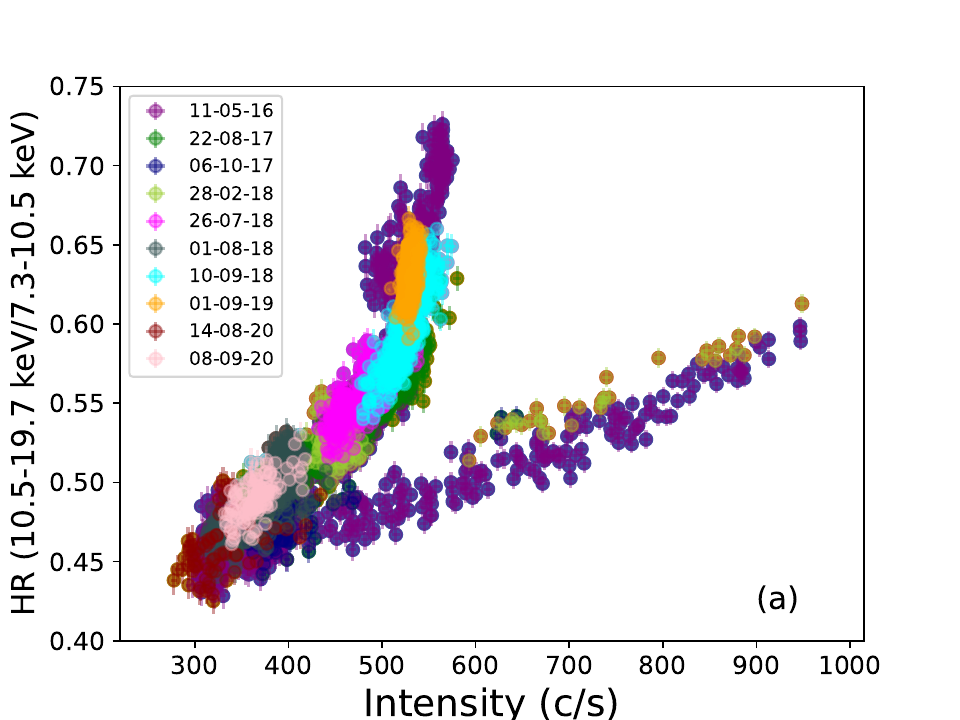}
\includegraphics[scale=0.59]{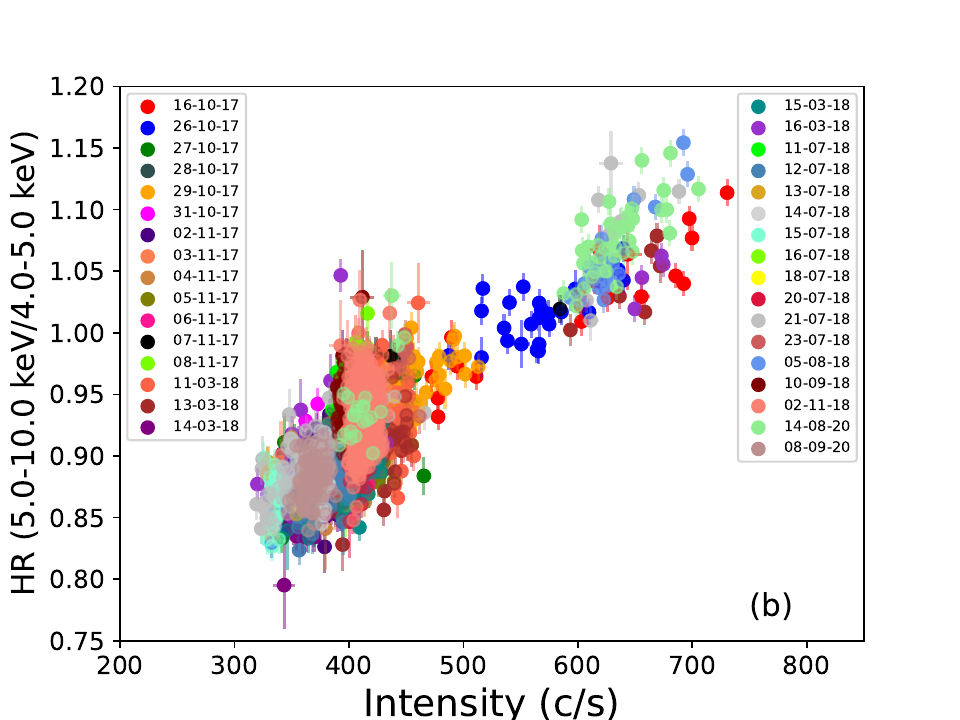}
\includegraphics[scale=0.59]{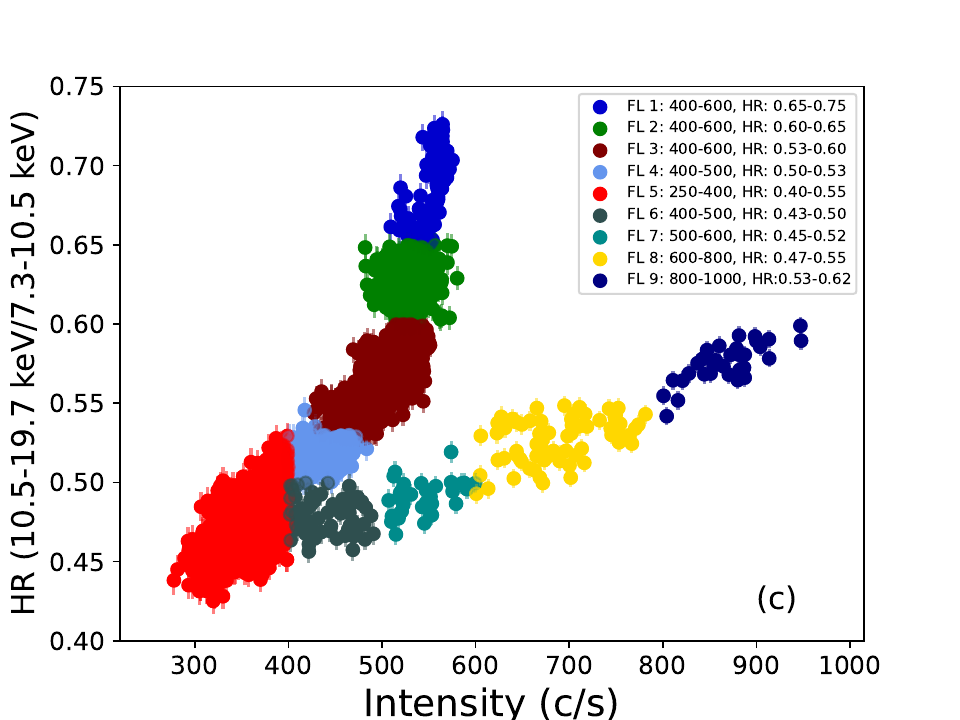}
\includegraphics[scale=0.59]{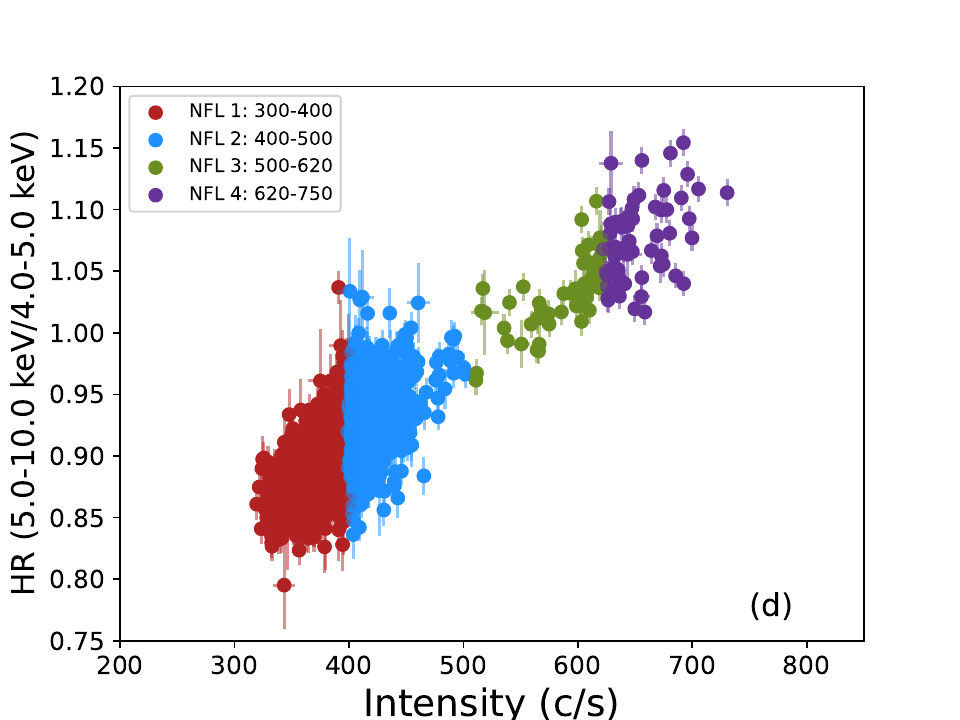}
\caption{\label{fig:obshid} Hardness-Intensity Diagram (HID) of GX 17+2 based on observations from (a) AstroSat and (b) NICER, and flux-resolved HID from (c) AstroSat and (d) NICER. In panels (a) and (b), different colors represent individual AstroSat and NICER observations, respectively. Panels (c) and (d) depict flux levels: FL 1 to 9 for AstroSat and NFL 1 to 4 for NICER, with segmentation as indicated in the legend. The Y-axis shows the hardness ratio (HR), while the X-axis represents the source intensity across the total energy range. Data points are binned to 32 s intervals. }
\end{figure*}

 AstroSat \citep{2014Singh} has observed the source GX 17+2 for eleven times up to the year 2020. We excluded the AstroSat data with observation ID G05\_142T01\_9000000610 of date August 18, 2018, because of a large angular offset in the observation, caused by incorrect pointing to that source during the observation. Figure~\ref{fig:obshid}(a) displays the source's HID using all the AstroSat observations considered, which is collectively tracing a portion of the Z-path. The  AstroSat observation log is presented in Table~\ref{tab:1}.\\
LAXPC has three alike proportional counters, LAXPC 10, 20, and 30, with a working energy range of 3--80 keV. Each LAXPC observation is reduced using the standard pipeline of  {\tt LAXPCsoftware}\footnote{\url{ http://astrosat-ssc.iucaa.in/laxpcData}} (latest version: Oct. 13, 2022). The spectrum, background spectrum, and responses are extracted following the typical procedure detailed in ASSC website\footnote{\url {http://astrosat-ssc.iucaa.in/uploads/threadsPageNew_SXT.html}}. However, LAXPC 10 and 30 experienced aberrant
gain changes and incurred gas leaks, respectively, from March 2018 onwards; thus, we took into account the LAXPC 20 spectrum for all observations. The topmost layer (L1) has been used in order to minimize the background (for more details, refer to \citet{2021Antia, 2019Beri, 2022Nath, 2024aBhattacherjee, 2024bBhattacherjee}). \\
SXT operates in the soft X-ray energy of 0.3--8.0 keV \citep{2016Singh, 2017Singh}. We used the refined level2 SXT data of different orbits as provided in {\sc Astrobrowse}\footnote{\url{https://astrobrowse.issdc.gov.in/astro_archive/archive/Home.jsp}}, then merged them into the final event file with the help of {\sc SXTMerger}\footnote{\url {https://www.tifr.res.in/~astrosat_sxt/index.html}} tool. A uniform annular region with inner and outer radius of 2 and 15 arcmin, respectively, was used for all the observations to eliminate the pile-up effect from the source counts.  {\tt XSELECT V2.5b} of {\tt HEASOFT (v. 6.33.1)} has been further utilized to extract the spectrum from the event file for the associated observations. The corresponding vignetting corrected ancillary response files (ARF) were extracted using {\tt sxtARFmodule}\footnote{\url{https://www.tifr.res.in/~astrosat_sxt/dataanalysis.html}} for the spectral analysis. The standard background spectrum ({\tt SkyBkg\_comb\_EL3p5\_Cl\_Rd16p0\_v01.pha}) and response ({\tt sxt\_pc\_mat\_g0to12.rmf}) of the instrument have been used as provided by the SXT team\footnote{\url{http://astrosat-ssc.iucaa.in/sxtData}}.

 NICER has observed the source thirty-seven times up to the year 2020 since its launch.  Figure~\ref{fig:obshid}(b) shows the source's HID using all the NICER observations together. The  NICER observation log is presented in Table~\ref{tab:1}.\\ X-ray timing instrument (XTI) is the primary instrument of  NICER, operational in the soft X-ray energy band of 0.2-12.0 keV. XTI is made up of an array of 56 silicon drift detectors (SSDs) and X-ray optics. Of this assembly, 52 are in operational condition \citep{2012Gendreau,2014Arzoumanian}. We used the  NICER CALDB version 20240206\footnote{\url{https://heasarc.gsfc.nasa.gov/docs/heasarc/caldb/nicer/}}.  NICER spectra have been generated using {\tt nicerl3-spect} pipeline, which generates the corresponding response and ancillary response files for the given customized good time interval data. For background estimation, we use the {\tt nibackgen3c50} model \citep{2022Remillard}, being one of the most luminous XRBs, the background contribution of the sources is not much significant. 
\section{Flux-resolved spectral analysis along the track} \label{sec:sa}
\subsection{AstroSat Analysis}\label{sec:aa}
To carry out the flux-resolved spectral analysis, we analyzed the HID created using the simultaneous LAXPC and SXT data of all the AstroSat observation, as shown in Figure~\ref{fig:obshid}(a). The HID is obtained using the hardness ratio (HR) of the X-ray photons in the energy range 10.5–19.7 keV and 7.3–10.5 keV \citep{2002Homan} and intensity in the 7.3–19.7 keV energy range. Using the LAXPCsoftware subroutine {\tt laxpc\_flux\_HR\_res}, the curve is segmented into nine flux levels (FLs): FL 1 to FL 9 (FL 1: 400-600 c/s, HR 0.65-0.75; FL 2: 400-600 c/s, HR: 0.60-0.65; FL 3: 400-600 c/s, HR: 0.53-0.60; FL 4: 400-500 c/s, HR: 0.50-0.53; FL 5: 250-400 c/s, HR: 0.40-0.55; FL 6: 400-500 c/s, HR: 0.43-0.50; FL 7: 500-600 c/s, HR: 0.45-0.52; FL 8: 600-800 c/s, HR: 0.47-0.55; FL 9: 800-1000 c/s, HR:0.53-0.62) as represented in Figure~\ref{fig:obshid}(c). For each of the FLs, we extracted the spectrum with their corresponding background spectra and response files for LAXPC and vignetting corrected ARFs for SXT. The spectrum was optimally grouped using the ftool {\tt ftgrouppha}. We did the LAXPC-SXT joint spectral analysis using {\tt XSPEC 12.14.0b} \citep{1996Arnaud} over an energy range of 1-20 keV (SXT: 1-7 keV; LAXPC: 4-20 keV) for all the observations belonging to a particular FL. Energies below 1 keV were not taken into account for any AstroSat observations due to uncertainties in the effective area and response of the SXT. In joint fitting, we incorporated a model systematic error of 3\% as suggested by AstroSat team\footnote{\url {https://www.tifr.res.in/~astrosat_sxt/dataana_up/readme_sxt_arf_data_analysis.txt}}.

\begin{figure}
\includegraphics[scale=0.59]{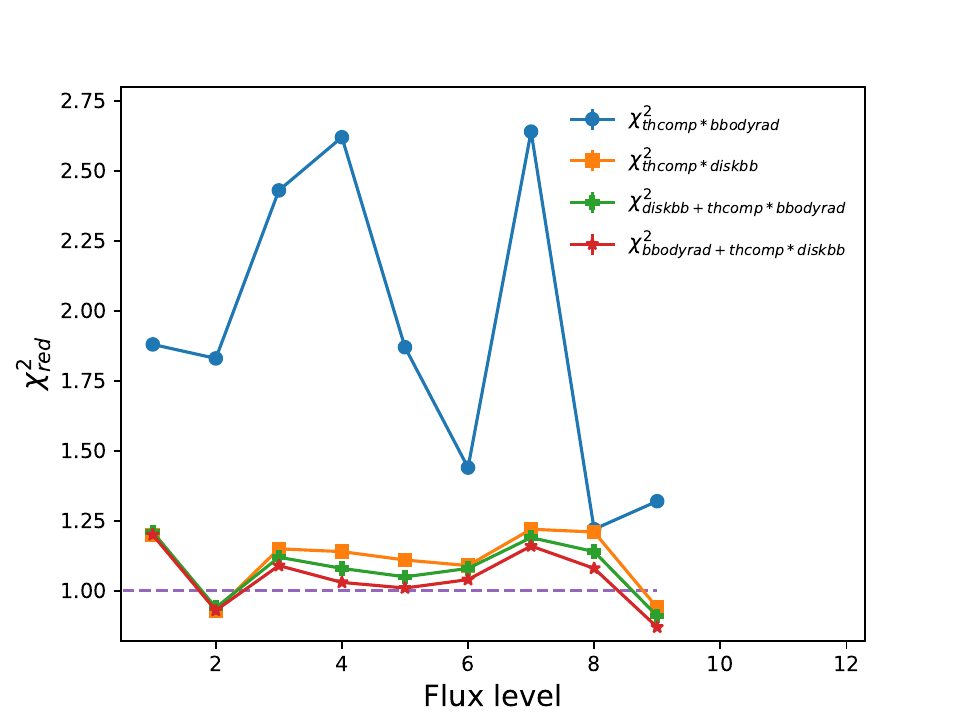}
\caption{\label{fig:rchi}{Comparison of $\chi^2_{\rm red}$ value for all flux levels using the four tested models as indicated in the legend for the observed spectra.}}
\end{figure}

\vskip 0.5cm
\begin{figure*}
\includegraphics[scale=0.32, angle=-90]{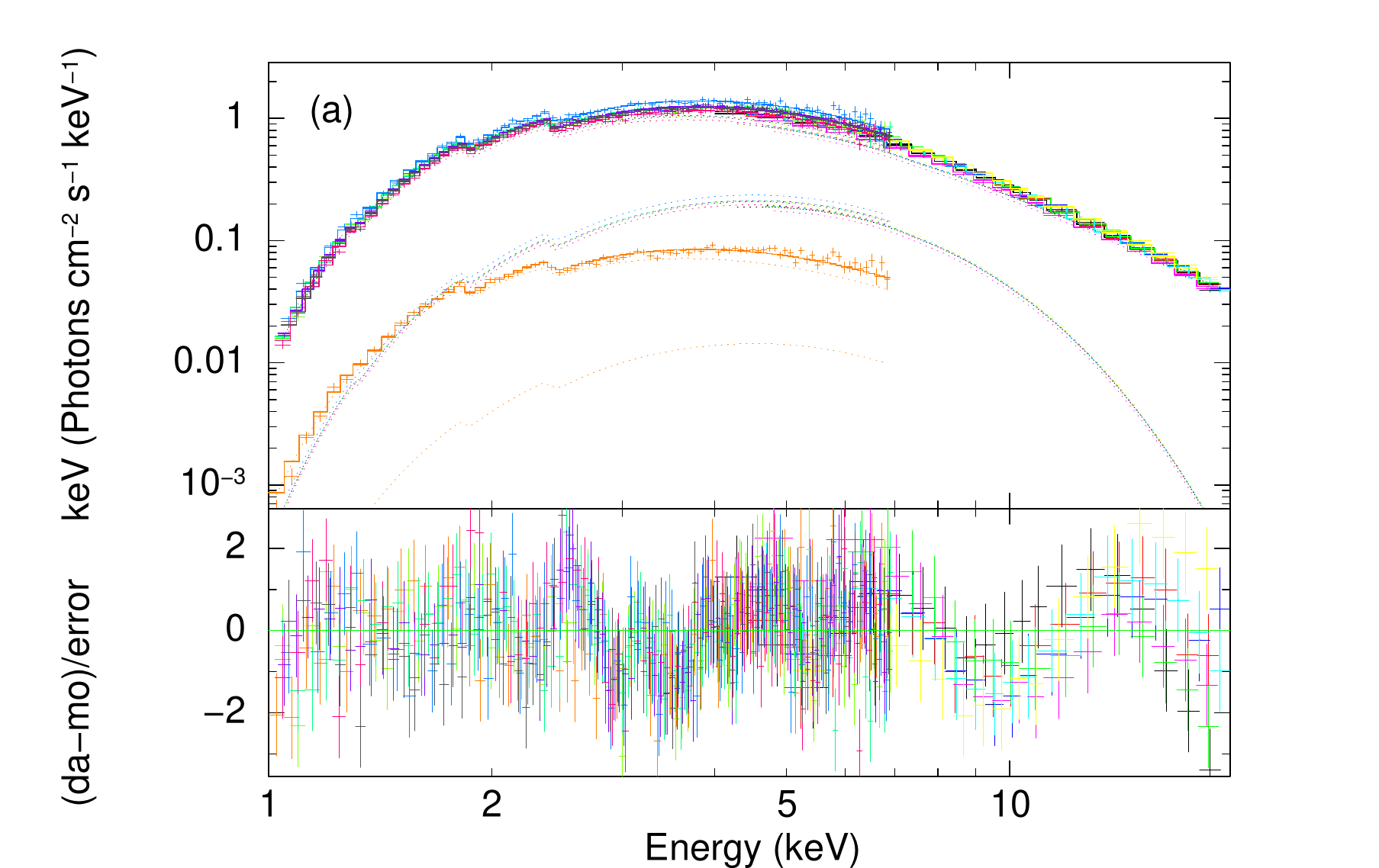}
\includegraphics[scale=0.32, angle=-90]{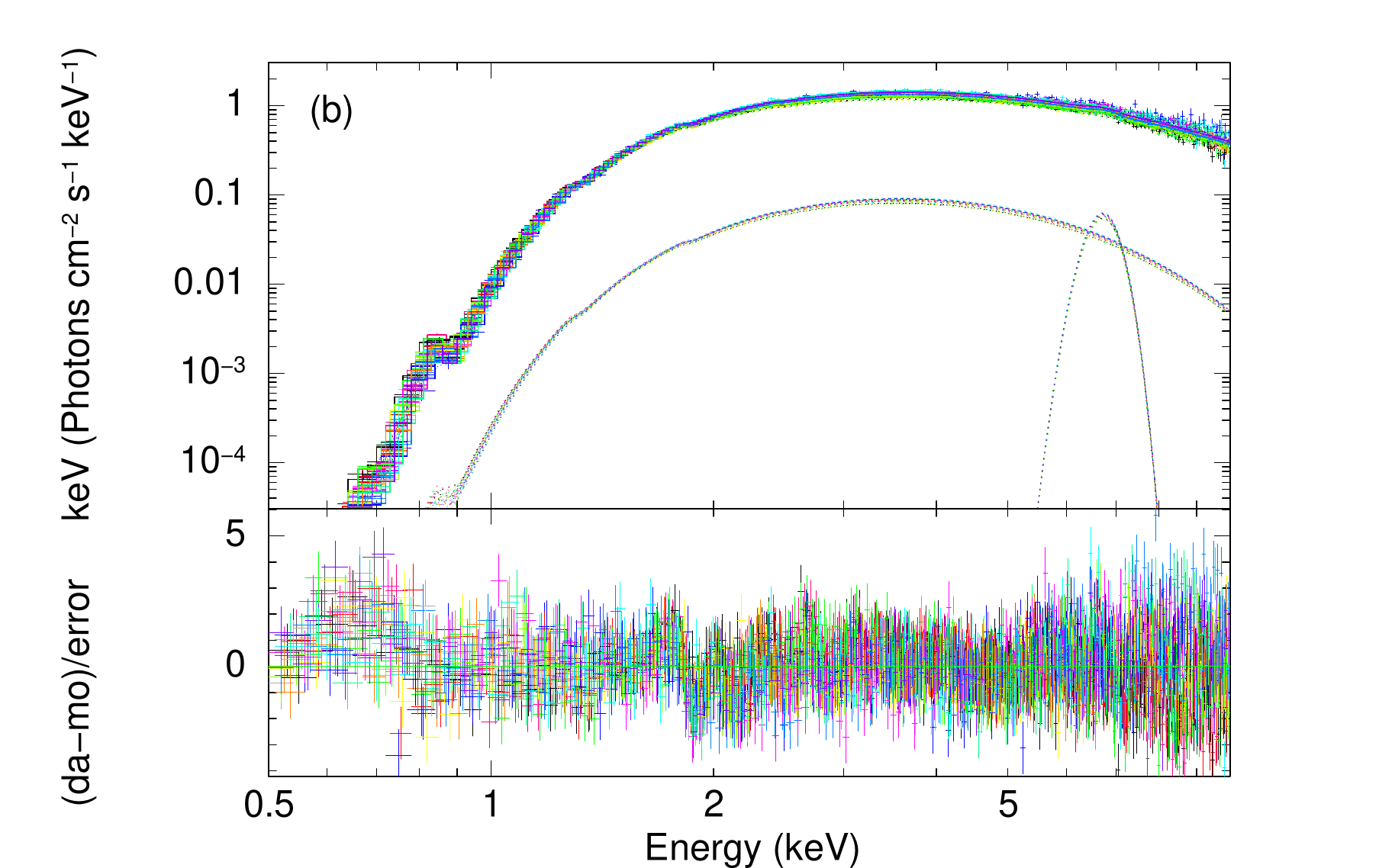}
\caption{\label{fig:specfit} The  best-fit  representative spectra for the model {\tt tbabs*(bbodyrad+thcomp*diskbb)} ; (a) flux level 5 from AstroSat, over the energy range of 1-20 keV, and (b)  flux level 1 from NICER, over the energy range
of 0.5-10.0 keV of the X-ray spectrum. NOTE: One SXT observation is in FW mode, leading to a relatively low SXT constant for that specific data group. }
\end{figure*}

\begin{table*}
	\centering
	\setlength{\tabcolsep}{3pt}
	\caption{Flux-resolved spectroscopy: Best-fit spectral parameters for the  AstroSat observations using the model combination {\tt tbabs*(bbodyrad+thcomp*diskbb)} .}
	\label{tab:tab3}
	\begin{tabular}{@{}cccccccccc@{}} 
		\hline\hline
Parameter  &  FL 1 & FL 2 &FL 3 &FL 4 & FL 5 & FL 6& FL 7 & FL 8 &FL 9 \\
	\hline
N$_{\rm H}$ ($10^{22} \rm{cm}^{-2})$ & $2.12^{+0.06}_{-0.06}$  & $2.11^{+0.03}_{-0.02}$  & $2.11^{+0.01}_{-0.01}$ &  $2.08^{+0.01}_{-0.01}$  & $2.09^{+0.01}_{-0.01}$ & $2.06^{+0.03}_{-0.03}$&$2.06^{+0.03}_{-0.02}$ & $2.16^{+0.04}_{-0.04}$&    $2.16^{+0.06}_{-0.05}$ \\\\
\hline
$\rm kT_{bb}$ (keV) &  $1.56^{+0.13}_{-0.24}$&$1.51^{+0.11}_{-0.19}$&$1.44^{+0.08}_{-0.10}$ &$1.43^{+0.07}_{-0.07}$& $1.37^{+0.06}_{-0.06}$& $1.38^{+0.07}_{-0.07}$&$1.56^{+0.14}_{-0.19}$&  $1.52^{+0.08}_{-0.08}$ &$1.54^{+0.14}_{-0.16}$\\\\
\hline
$\rm kT_e$ (keV) & $3.14^{+0.23}_{-0.14}$& $3.15^{+0.15}_{-0.09}$  &$3.07^{+0.06}_{-0.05}$&$3.17^{+0.08}_{-0.07}$&  $3.11^{+0.07}_{-0.06}$&$3.02^{+0.11}_{-0.09}$&$3.03^{+0.23}_{-0.13}$&$3.46^{+0.34}_{-0.24}$  &  $3.17^{+0.35}_{-0.21}$\\\\

f$_s$&$0.12^{+0.02}_{-0.02}$&$0.10^{+0.01}_{-0.01}$ & $0.08^{+0.01}_{-0.01}$& $0.06^{+0.01}_{-0.01}$&$0.05^{+0.01}_{-0.01}$ &$0.05^{+0.01}_{-0.01}$  &  $0.07^{+0.01}_{-0.01}$&$0.05^{+0.01}_{-0.01}$ &  $0.09^{+0.03}_{-0.03}$\\\\

$\Gamma$&$<1.85$ &$<1.56$&$<1.25$&$<1.25$&$<1.22$&$<1.55$&$<1.63$&$<2.33$&$<2.44$\\\\

$\tau$&$>5.24$ &$>12.97$&$>25.66$&$>22.45$&$>27.05$&$>15.57$&$>12.12$&$>6.78$&$>4.44$\\\\

\hline
T$_{\rm in}$ (keV) &    $1.41^{+0.17}_{-0.09}$ &$1.54^{+0.15}_{-0.07}$&   $1.68^{+0.08}_{-0.06}$& $1.74^{+0.06}_{-0.06}$ &$1.66^{+0.05}_{-0.05}$ &   $1.71^{+0.04}_{-0.04}$  & $1.75^{+0.18}_{-0.10}$&    $2.09^{+0.09}_{-0.09}$  & $2.11^{+0.19}_{-0.17}$\\\\

N$_{\mathrm{dbb}}$&$153.55^{+40.34}_{-48.09}$&$118.30^{+21.11}_{-28.74}$ & $92.35^{+12.00}_{-12.74}$& $81.96^{+9.41}_{-8.88}$&$86.06^{+9.38}_{-8.18}$&$94.06^{+11.01}_{-9.87}$&$88.14^{+18.41}_{-22.35}$&$55.74^{+10.68}_{-8.68}$&$57.38^{+20.50}_{-15.53}$\\

R$_{\rm{in}}$ \text{ (km)} &51.44$^{+6.36}_{-8.81}$ & 45.15$^{+3.86}_{-5.87}$ & 39.89$^{+2.51}_{-2.85}$& 37.58$^{+2.10}_{-2.09}$ &38.51$^{+2.04}_{-1.88}$ &40.26$^{+2.29}_{-2.17}$ &38.97$^{+3.88}_{-5.30}$ &30.99$^{+2.84}_{-2.52}$& 31.44$^{+5.19}_{-4.59}$ \\

\hline
$\chi^2_{\rm red}$&1.20&0.93&1.09&1.04&1.01&1.04&1.16&1.08&0.87\\
\hline
$\dot{M}$ $\times$10$^{18} (gm/s)$  
&2.97$^{+0.19}_{-0.15}$ &2.93$^{+0.23}_{-0.44}$  &2.85$^{+0.17}_{-0.22}$  &2.70$^{+0.15}_{-0.14}$
&2.46$^{+0.10}_{-0.13}$ &3.07$^{+0.14}_{-0.10}$  &3.11$^{+0.28}_{-0.50}$ &3.16$^{+0.04}_{-0.03}$ &3.44$^{+0.43}_{-0.48}$\\  

\hline
Flux$_{\rm tot}$$\times$10$^{-8}$ & 2.29$^{+0.05}_{-0.05}$& 2.30$^{+0.04}_{-0.03}$ & 2.26$^{+0.03}_{-0.03}$ & 2.18$^{+0.03}_{-0.03}$ & 1.87$^{+0.03}_{-0.03}$&2.24$^{+0.04}_{-0.04}$ &2.50$^{+0.04}_{-0.04}$& 2.98$^{+0.05}_{-0.05}$ & 3.32$^{+0.08}_{-0.08}$ \\
(erg cm$^{-2}$ s${-1}$)&&&&&&&&&\\
\hline
Compton flux$\times$10$^{-9}$ &9.94$^{+0.94}_{-1.53} $& 8.55$^{+0.82}_{-1.45} $&6.61$^{+0.74}_{-0.90} $&5.64$^{+0.67}_{-0.77}$ &
4.48$^{+0.57}_{-0.59}$& 5.22$^{+0.95}_{-1.05} $& 7.18$^{+1.26}_{-1.97} $ &6.91$^{+0.76}_{-0.75} $&
8.58$^{+1.42}_{-1.45}$\\

(erg cm$^{-2}$ s${-1}$)&&&&&&&&&\\
\hline
Disk flux$\times$10$^{-8}$ & 1.31$^{+0.14}_{-0.08}$&1.45$^{+0.14}_{-0.07}$&1.60$^{+0.09}_{-0.07}$ & 1.61$^{+0.07}_{-0.06}$ &1.43$^{+0.05}_{-0.05}$ &1.72$^{+0.10}_{-0.09}$ &1.78$^{+0.19}_{-0.12}$ &$^{+0.06}_{-0.06}$ &2.46$^{+0.12}_{-0.12}$\\
(erg cm$^{-2}$ s${-1}$)&&&&&&&&&\\
\hline

\small{\text{BB flux}}$\times$10$^{-9}$
&$3.78^{+0.94}_{-0.91}$ 
& $3.35^{+0.49}_{-0.51}$ 
& $2.95^{+0.38}_{-0.37}$ 
& $3.30^{+0.44}_{-0.43}$ 
&$3.16^{+0.46}_{-0.45}$ 
& $4.07^{+1.00}_{-1.02}$ 
& $3.83^{+0.77}_{-0.75}$ 
& $7.03^{+1.43}_{-1.49}$ 
& $5.26^{+2.07}_{-2.16}$ \\
\small{(erg cm$^{-2}$ s${-1}$)}&&&&\\

\hline
L$_{\rm T}$$\times$10$^{38}$ &$4.65^{+0.11}_{-0.10}$ &$4.66^{+0.07}_{-0.07}$& $4.57^{+0.06}_{-0.06}$ & $4.40^{+0.06}_{-0.06}$ & $3.79^{+0.05}_{-0.05}$&$4.54^{+0.08}_{-0.08}$& $5.05^{+0.09}_{-0.08}$&$6.02^{+0.10}_{-0.10}$& $6.71^{+0.16}_{-0.15}$\\
(erg s$^{-1}$)&&&&&&&&&\\
\hline
Efficiency ($\eta$) & 0.17$^{+0.01}_{-0.01}$ &0.18$^{+0.01}_{-0.01}$&  0.18$^{+0.01}_{-0.01}$ & 0.18$^{+0.01}_{-0.01}$ &  0.17$^{+0.01}_{-0.01}$ & 0.16$^{+0.01}_{-0.01}$ & 0.18$^{+0.03}_{-0.02}$&  0.21$^{+0.01}_{-0.02}$ &  0.21$^{+0.03}_{-0.03}$ \\
\hline\hline
\multicolumn{10}{p{18cm}}{Note: Optical depth ($\tau$) is fixed at 30 for all FLs.}
\end{tabular}		
\end{table*}
\begin{figure*}
\center
\includegraphics[scale=0.52, angle=0]{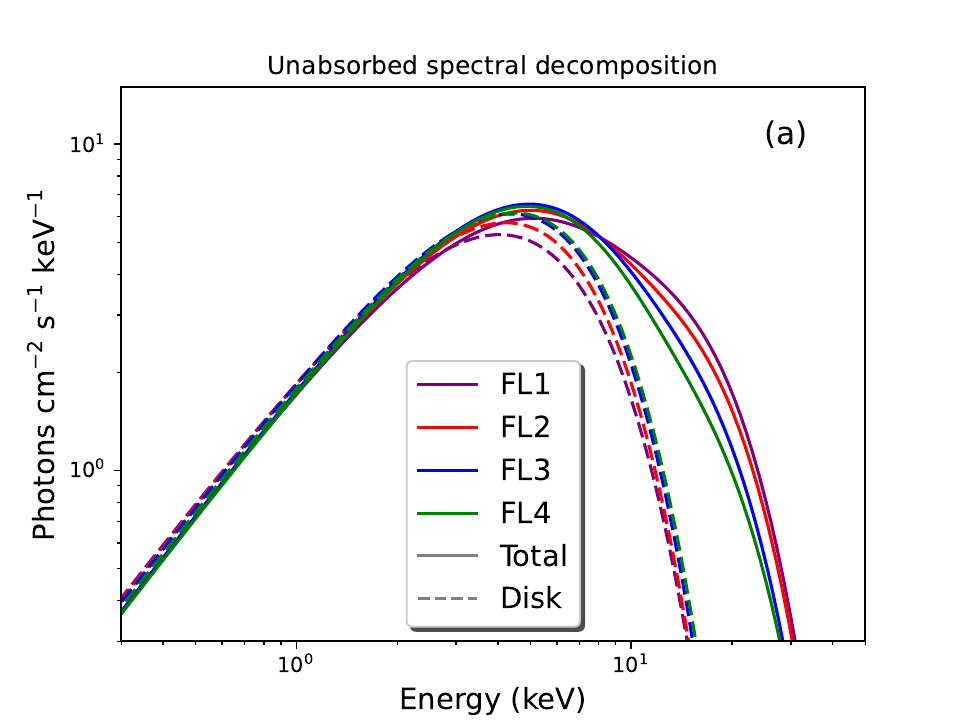}
\includegraphics[scale=0.52, angle=0]{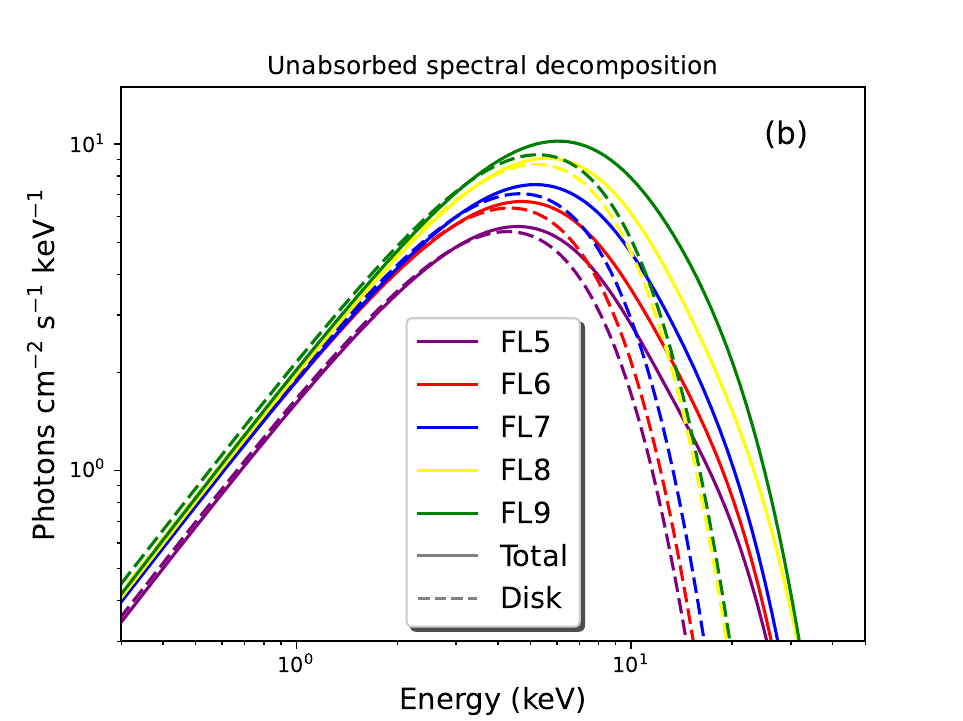}
\caption{\label{fig:model} {The best-fit model {\tt tbabs*(bbodyrad+thcomp*diskbb)} for the best-fit parameter values for (a) NB and  (b) FB. It illustrates the relative contributions of each spectral component in NB and FB, showing the increasing dominance of the disk component (`- - -') as the spectrum transitions from harder to softer states along each branch. The total model contribution (Compton+disk, `\textemdash') is stronger in NB than FB reflecting the significant contribution of Comptonization in NB. The absorption component is fixed to 0, to get the unabsorbed spectral decomposition.  }}
\end{figure*} 

\begin{figure*}
\includegraphics[scale=0.36]{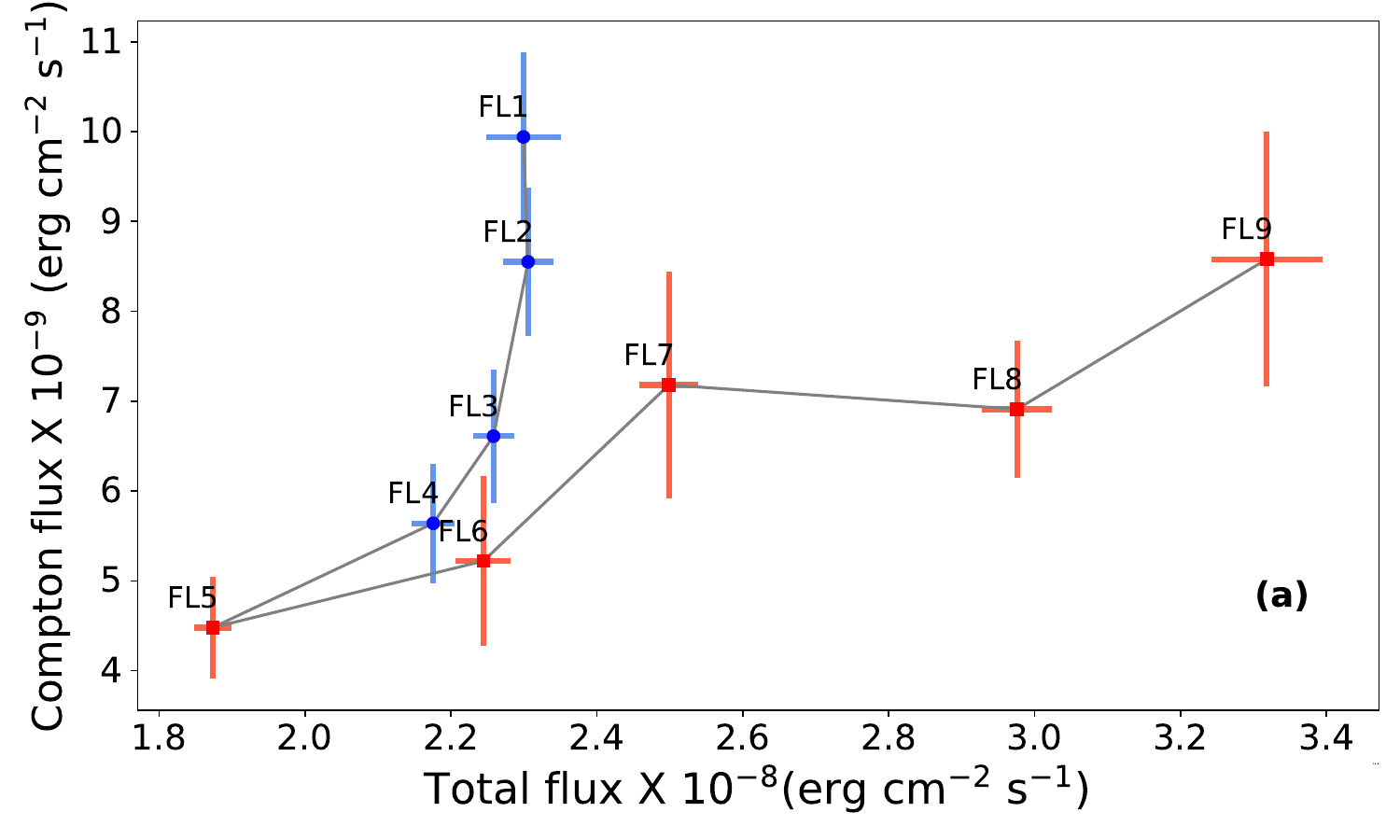}
\includegraphics[scale=0.36]{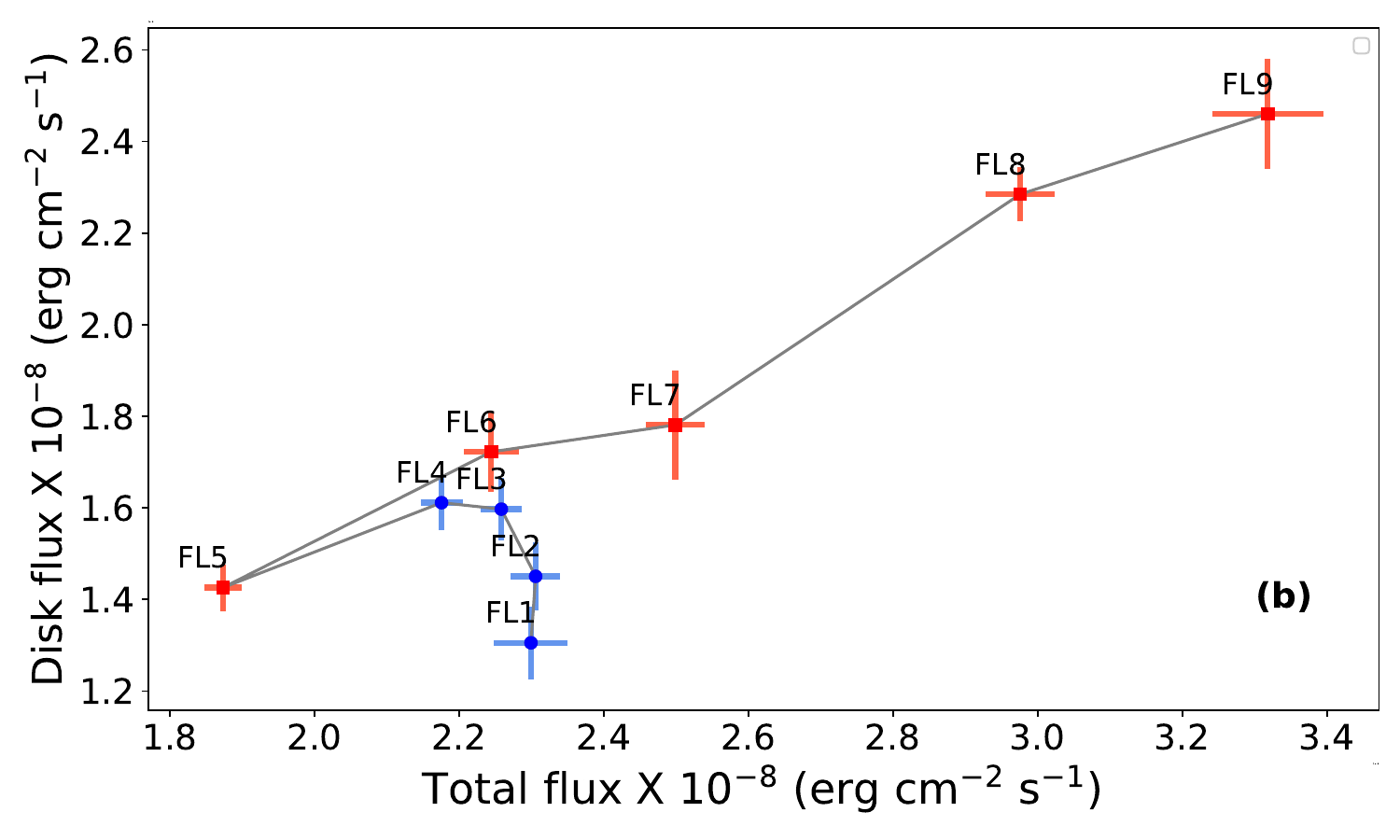}
\caption{\label{fig:tf} Variation of (a) Compton flux, and (b) Disk flux relative to the total flux, respectively from the AstroSat analysis using the model combonation {\tt tbabs*(bbodyrad+thcomp*diskbb)}, illustrating a Z-pattern in the traced path. In the figure, circular and square data points represents the parameter along the NB and FB branch, respectively.}
\end{figure*} 

We started fitting with the simple model combination, considering the NS blackbody surface to be the source of seed photons that are being Comptonized, i.e., tbabs*(thcomp*bbodyrad). But while doing so, in some segments the fit was not good with $\chi^2_{\rm red}$ $\sim$ 2 with a very high normalization value resulting in an unphysical blackbody radius value. It reveals that the signature of blackbody emission is not strong here. Then we replaced the blackbody emission with the disk emission component, or {\tt diskbb} \citep{1984Mitsuda, 1986Makishima}. This results in a better fit in comparison to the previous assumption, with $\chi^2_{\rm red}$ $\sim$ 1 in all the FLs. 
 Then we tested the combination of both disk and blackbody surface emisison components as follows: {\tt diskbb+thcomp*bbodyrad} and {\tt bbodyrad+thcomp*diskbb}.  
In the model combination {\tt bbodyrad+thcomp*diskbb}, this resulted in a slight improvement in the fit, while the variation of the key parameters remained mostly unchanged, yielding all parameters within acceptable values. Next we consider the other possible combination, {\tt diskbb+thcomp*bbodyrad}, we found similar $\chi^2_{\rm red}$ with comparable parametric variation. In this work we primarily show the results from the model combination {\tt bbodyrad+thcomp*diskbb} and while we also discuss the interpretation from the combination {\tt diskbb+thcomp*bbodyrad} in the Section~\ref{sec:alt}. The comparison of the $\chi^2_{\rm red}$ for all FLs along the track, using all model combinations, is shown in Figure~\ref{fig:rchi}. The Interstellar medium model,{\tt tbabs} is used to account for the Galactic absorption \citep{2000Wilms}. We used two edge components at low energies, $\sim$2.42 keV and  $\sim$1.86 keV, associated with Au and Si, respectively \citep{2017Singh}. The model {\tt bbodyrad}, which takes care of the blackbody surface emission, has two parameters: the blackbody temperature (kT$_{\rm bb}$) and the blackbody normalization (N$_{\rm bb}$). The normalization parameter N$_{\rm bb}$ is related to the blackbody radius (R$_{\rm bb}$) of the NS through the equation: N$_{\rm bb}$= R$_{\rm bb}^2$/D$_{\rm 10}^2$, where R$_{\rm bb}$ is the blackbody radius in km and D$_{\rm 10}$ is the distance to the source in units of 10 kpc. While fitting we fixed the N$_{\rm bb}$ to 10 km, which is typical NS blackbody radius. The Comptonization model component {\tt ThComp} \citep{2020Zdziarski} accounts for thermal Comptonization and includes three other important parameters: electron temperature ($\rm kT_{\rm e}$),  which represents the hot coronal temperature; asymptotic power-law index ($\Gamma$), which represents the spectral index; and the covering fraction (f$\rm_s$), which can vary from 0 to 1 depending on the observed seed photons that are being Comptonized or scattered by the hot coronal electrons. For instance, if it is set to 0, it refers that only original seed photons will be seen, while for 1 it indicates all the seed photons are getting Comptonized.

The {\tt diskbb} model accounts for emission from the accretion disk, providing information about the inner-disk temperature  (T$_{\rm in}$) and inner-disk radius (R$_{\rm in}$) through the disk normalization parameter (N$_{\rm dbb}$). Table~\ref{tab:tab3} frames the best-fit spectral parameters with a 90\% confidence level. As shown in the  Table~\ref{tab:tab3}, the lower bound of $\Gamma$ parameter could not be constrained. For kT$_e$ $\ll$ m$\rm_e$c$^2$, the optical depth ($\tau$) is related to $\Gamma$ by the relation \citep{2020Zdziarski}:
\begin{equation}
  \Gamma(\tau) = \sqrt{\frac{9}{4} + \frac{1}{\theta \, \tau (a_1 + a_2 \tau)}} - \frac{1}{2}
  \label{eq:gam}
\end{equation}
\noindent
where $\theta$ = kT$_e$/m$_e$c$^ 2$, a$_1$ = 1.2 and a$_2$ = 0.25. Since $\tau$ is more physical and is related to $\Gamma$, we converted $\Gamma$ to $\tau$ during the spectral fitting. As evident from the values, since the upper bound of $\tau$ could not be constrained, we set its value to $\sim$30, based on the maximum lower limit of $\sim$27, and proceeded with the spectral analysis. While Figure~\ref{fig:specfit} (a) is a representative best-fit spectrum of FL 5, where the different colors represent the spectra of the different observations within the corresponding FL. Figure~\ref{fig:model} represents the relative contribution of the thermal and non-thermal components in the unabsorbed spectral decomposition plot along the NB (Figure~\ref{fig:model} (a)) and FB (Figure~\ref{fig:model} (b)) in an attempt to apprehend their contribution in the formation of Z-track. From Figure~\ref{fig:model} (a), it is clear that most of the variation in the NB is evident only above 10 keV which is beyond NICER. In other words the NFL 2 segment of NICER HID may contain the times when the source was moving along NB and has been clubbed into one state for NICER.

\subsection{NICER Analysis}\label{sec:na}

In this section, we analyzed the source along the NICER HID, where the HR of the X-ray photons is defined in the energy range 4.0–10.0/0.5–4.0 keV and intensity in the energy range of 0.5–10.0 keV. The curve is segmented into four NICER FLs (NFLs): NFL 1 to NFL 4 (NFL 1: 300-400 counts/s, NFL 2: 400-500 counts/s, NFL 3: 500-620 counts/s, NF4: 620-750 counts/s); refer to Figure~\ref{fig:obshid}(d). For all the NFLs, we extracted the corresponding spectrum along with their ARFs and response files, which we used for the spectral fitting. The spectrum binning and required systematics are applied by default when generating the spectrum using {\tt nicerl3-spect}\footnote{\url{ https://heasarc.gsfc.nasa.gov/lheasoft/help/nicerl3-spect.html}}.

\begin{table}
	\centering
	\setlength{\tabcolsep}{3.0 pt}
	\caption{Flux-resolved spectroscopy: Best-fit spectral parameters for the  NICER observations using the model combination {\tt tbabs*(bbodyrad+thcomp*diskbb)} }
	\label{tab:tab4}
	\begin{tabular}{@{}ccccc@{}} 
		\hline\hline
\small{Parameter} & \small{NFL 1} & \small{NFL 2}& \small{NFL 3} &\small{ NFL 4 }\\ 
\hline
\small{kT$_{\mathrm{bb}}$ (keV)}&   $1.08^{+0.06}_{-0.05}$ &$1.26^{+0.11}_{-0.09}$ & $2.01^{+0.04}_{-0.09}$  &$2.16^{+0.06}_{-0.30}$ \\ 
\hline
\small{f$_{\mathrm{s}}$} &  $0.04^{+0.01}_{-0.01}$&$0.08^{+0.01}_{-0.01}$&$<$0.03 &$<$0.11 \\ 
\hline
\small{T$_{\mathrm{in}}$ (keV)}&  $2.03^{+0.04}_{-0.05}$ & $1.97^{+0.07}_{-0.09}$ & $2.19^{+0.09}_{-0.07}$&   $2.20^{+0.07}_{-0.06}$\\ 
\small{R$_{\mathrm{in}}$ (km)} & $28.55^{+0.78}_{-0.81}$ & $33.79^{+2.57}_{-1.87}$ & $26.97^{+ 1.49}_{- 1.52}$& $28.05^{+1.64}_{-1.47}$ \\ 
\hline
$\chi^2_{\rm red}$ & $1.15$ & $1.31$ &$1.01$ & $0.96$\\ 
\hline
$\dot{M}$ $\times$10$^{18} (gm/s)$  & $2.20^{+0.07}_{-0.07}$ & $3.28^{+0.26}_{-0.20}$&$2.54^{+0.19}_{-0.15}$&$2.88^{+0.37}_{-0.15}$ \\
\hline
        Flux$_{\rm tot}$ & $1.96^{+0.12}_{-0.12}$&$2.77^{+0.25}_{-0.25}$ &$3.15^{+0.36}_{-0.24}$ & $3.79^{+1.16}_{-0.44}$ \\
\scriptsize{$\times$10$^{-8}$ (erg cm$^{-2}$ s${-1}$)}&&&&\\
\hline
\small{\text{Comp. flux}} &$2.41^{+0.35}_{-0.39}$ & $5.98^{+0.89}_{-1.12}$& $10.5^{+0.72}_{-0.74}$ & $14.9^{+3.28}_{-0.86}$\\
\scriptsize{$\times$10$^{-9}$}\scriptsize{(erg cm$^{-2}$ s${-1}$)}&&&&\\
\hline
\small{\text{Disk flux}} &$1.72^{+0.03}_{-0.04}$& $2.17^{+0.09}_{-0.11}$ & $2.10^{+0.06}_{-0.07}$ &$2.30^{+0.31}_{-0.07}$ \\

\scriptsize{$\times$10$^{-8}$}\scriptsize{(erg cm$^{-2}$ s${-1}$)}&&&&\\
\hline
\small{\text{BB flux}} 
& $0.95^{+0.12}_{-0.12}$
& $1.90^{+0.17}_{-0.17}$ 
& $10.49^{+0.29}_{-1.37}$ 
& $12.62^{+1.28}_{-4.48}$ \\
\scriptsize{$\times$10$^{-9}$}\scriptsize{(erg cm$^{-2}$ s${-1}$)}&&&&\\

\hline\hline
\end{tabular}		
\end{table}

In this part of the study, we used the same spectral model component for the flux-resolved analysis to enable comparison with the results obtained from the AstroSat observation. Due to the high energy resolution and good sensitivity of NICER, it enables the detection of the iron emission line at $\sim$6.7 keV in the spectrum. In the lower energies, we added absorption energies, at $\sim$1.84 keV and $\sim$ 0.87 keV associated with Si K-edge and Ne K-edge, respectively\footnote{\url{ https://heasarc.gsfc.nasa.gov/docs/nicer/data_analysis/workshops/NICER-CalStatus-Markwardt-2021.pdf}}. While fitting, the kT$_e$ parameter was not well constrained since NICER is operational in soft energy. So taking reference from the AstroSat analysis, we fixed its value at $\sim$3 keV. The best-fit parameters are represented in Table~\ref{tab:tab4} which are in 90\% confidence level.  A representative best-fit spectrum of NFL 1 is represented in Figure~\ref{fig:specfit} (b); the different colors represent the spectra of the different observations within the corresponding NFL. 

Using the best-fit parameter from the AstroSat and NICER analysis, we computed the total bolometric unabsorbed flux (Flux$_{\rm tot}$) in the extended energy range of 0.01 to 100 keV, the disk flux, blackbody flux and Compton flux, using the {\tt XSPEC} model {\tt cflux}. The Compton flux results from subtracting the disk flux from the total bolometric unabsorbed flux. We also estimated the inner-disk radius (R$_{\rm in}$), $\dot{M}$, disk luminosity (L$_{\rm D}$) of the source along the different flux intensities. To calculate the R$_{\rm in}$ we use, \begin{equation}\label{eqn1} R_{\rm in}= \kappa^2(N_{\rm dbb}/cos\theta)^{1/2}*D_{10}\end{equation} where, D$_{10}$ is the distance to the source in units of 10 kpc which is 1.3 kpc for this source \citep{2008Galloway} and N$_{\rm dbb}$ is the best-fit disk normalization. We adopted the color-corrected factor ($\kappa$) to be 1.7 \citep{1995Shimura}, and the inclination of the source to be 35$^{\circ}$ \citep{2010Cackett, 2017Ludlam} throughout the calculation. The estimated R$_{\rm in}$ of the source ranges between $\sim$ 25--50 km combining the AstroSat and NICER analysis.\\

\begin{figure}
	\centering
	\includegraphics[height=20cm, width=8.5cm]{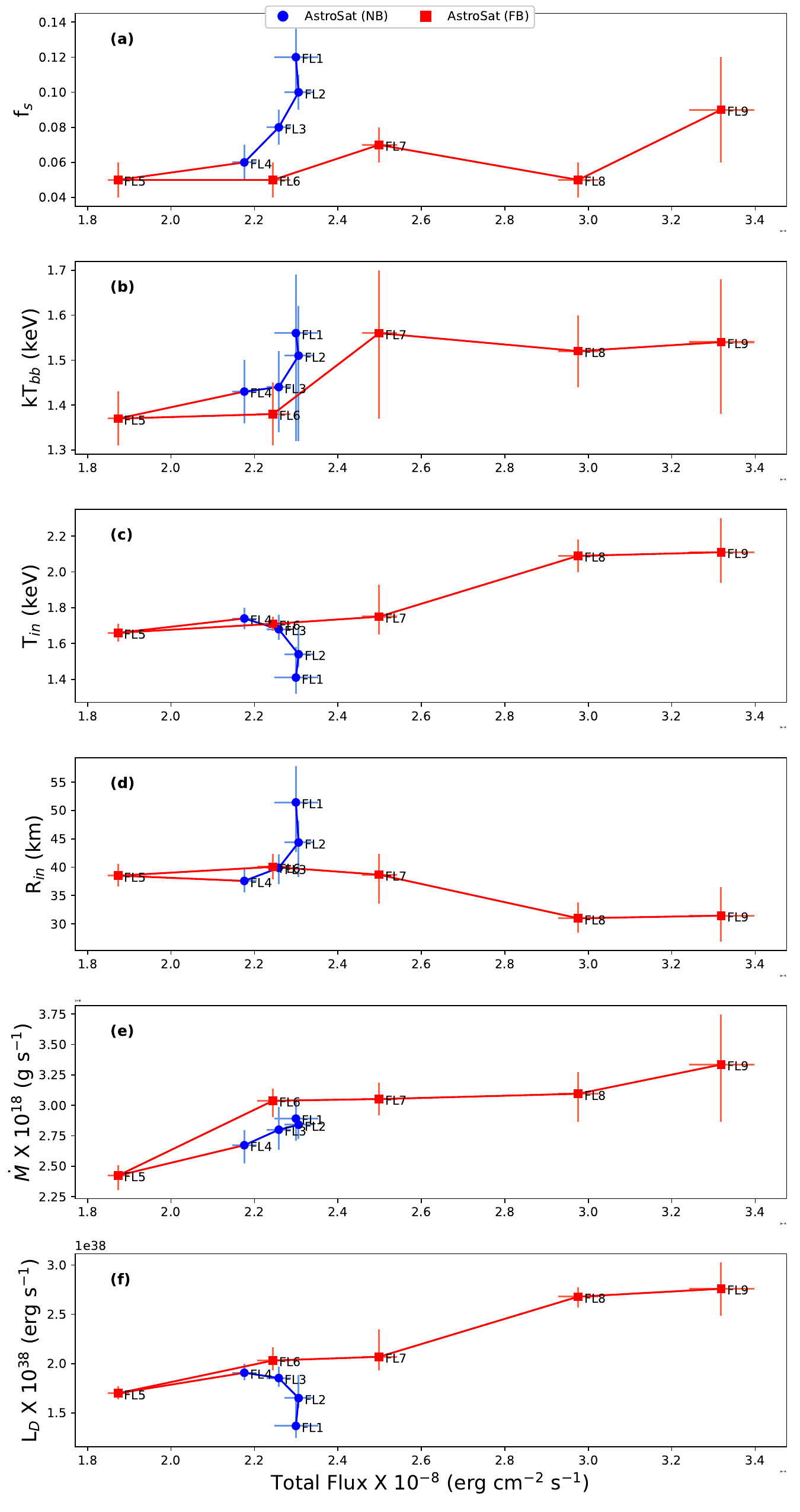}
    \caption{Variation in the source's spectral characteristics, obtained from flux-resolved spectroscopy using the model combination {\tt tbabs*(bbodyrad+thcomp*diskbb)}, relative to the total unabsorbed flux across different FL. The circular and square data points represent the source's properties across FL 1 to 4 (NB) and FL 5 to 9 (FB), respectively as observed by AstroSat. The plot reveals distinct change in behavior of the source from NB to the FB. The plot parameters are organized as follows: (a) covering fraction, f$_s$; (b) blackbody temperature, kT$_{bb}$; (c) inner-disk temperature, T$_{\rm in}$ in keV; (d) inner-disk radius, R$_{\rm in}$ in km; (e) mass-accretion rate, $\dot{M}$ in gm/s; (f) disk luminosity, L $_{\rm D}$ in erg s$^{-1}$, respectively. }
    \label{fig:as_vartf}
\end{figure}
Whereas, we estimated $\dot{M}$ using the equation 
\begin{equation} 
\label{eqn2}
T_{\text{in}} = \kappa\left(\frac{3GM\Dot{M}}{8\pi R_{\text{in}}^3 \sigma}\right)^{1/4} 
\end{equation} \citep{1984Mitsuda}
where M represents the typical mass of a NS (1.4$M_\odot$) , and $\sigma$ is the Stefan-Boltzmann constant. {\it G} is the gravitational constant, while T$\rm_{in}$ and R$\rm_{in}$ correspond to the inner-disk temperature and radius, respectively. Thus, both $R_{\rm in} = A N_{\rm dbb}^{1/2}$ and the accretion rate $\dot M = B A^3T_{\rm in}^4 N_{\rm dbb}^{3/2}$ can be obtained from the best fit values of the parameters $R_{\rm in}$ and $N_{\rm dbb}$, since $A$ and $B$ depend only on physical constants and  assumed values of distance, inclination angle and the color factor. However, it is not straightforward to compute the error $\dot M$, since in general the errors on  $T_{\rm in}$ and $N_{\rm dbb}$ are correlated. To circumvent this issue, we introduce a dummy variable to represent $\dot M$ in the {\tt XSPEC} environment. Then, since {\tt XSPEC} allows a parameter to be tied to other parameters in form of an expression, the parameter $T_{\rm in}$ can be tied to $\dot M$ and $N_{\rm dbb}$ with the expression $T_{\rm in}= (A^3B)^{-1/4}N_{\rm dbb}^{-3/8}{\dot M}^{1/4}$. In practice, the dummy variable $\dot M$ is introduced by adding a power-law with zero normalization (so that it does not contribute to the model spectrum) and identifying the index value as $\dot M$. With this setup, the error on $\dot M$ can be obtained directly from the {\tt XSPEC} error command. We estimated $\dot{M}$ of the source to be $\sim$ 2 -- 3.5 $\times$ 10$^{18}$ gm/s. \\ Additionally, we estimate L$_{D}$ of the source along with uncertainty, using the relation, \begin{equation}
\text{L}_{\text{D}} = \frac{2\pi d^2 F_D}{\cos\theta}
\end{equation}
where,  $\rm cos\theta$ is the inclination angle of the source, d is the distance of the source in cm and $F_D$ is the disk flux. In Section~\ref{sec:r}, we discuss the behavior of the flux with the spectral parameters along the track in detail. Table~\ref{tab:tab3} and \ref{tab:tab4} contains all the estimated values from AstroSat and NICER analysis, along with associated uncertainties.

\section{Results} \label{sec:r}

\subsection{Flux-correlation of the GX 17+2 }\label{sec:fluxcor}

In this section, we study the co-relation of the source properties along the different flux intensities using the primary model combination {\tt tbabs*(bbodyrad+thcomp*diskbb)}. Initially, we studied the variation of the Flux$_{\rm tot}$, disk flux, and the Compton flux along the independent FLs. Figure~\ref{fig:tf} shows the behavior of the Compton flux and the disk flux with the Flux$_{\rm tot}$ along the FLs as obtained from the AstroSat analysis. It is well evidenced from Figure~\ref{fig:tf} that the source is tracing towards the soft state and the Flux$_{\rm tot}$ is  primarily increasing after FL 5 by $\sim$ 77\%, indicating variation in $\dot{M}$. Because of the source's sudden change of properties from F5, we interpret the state as the transition state between NB and FB or SA. In Figure~\ref{fig:tf}(a), we can see there is a significant decrease in Compton flux with FLs till FL 4, then the Compton flux becomes constant or within the error bars, and the total flux increases dominantly, tracing an impression like the source's HID (Figure~\ref{fig:obshid}(a)). While in the case of disk flux vs Flux$_{\rm tot}$ in Figure~\ref{fig:tf}(b), we can observe the behavior satisfying the assumption of source tracing the soft state where the disk dominates the Comptonisation.

\begin{figure}
	\centering
	\includegraphics[height=6cm, width=8cm]{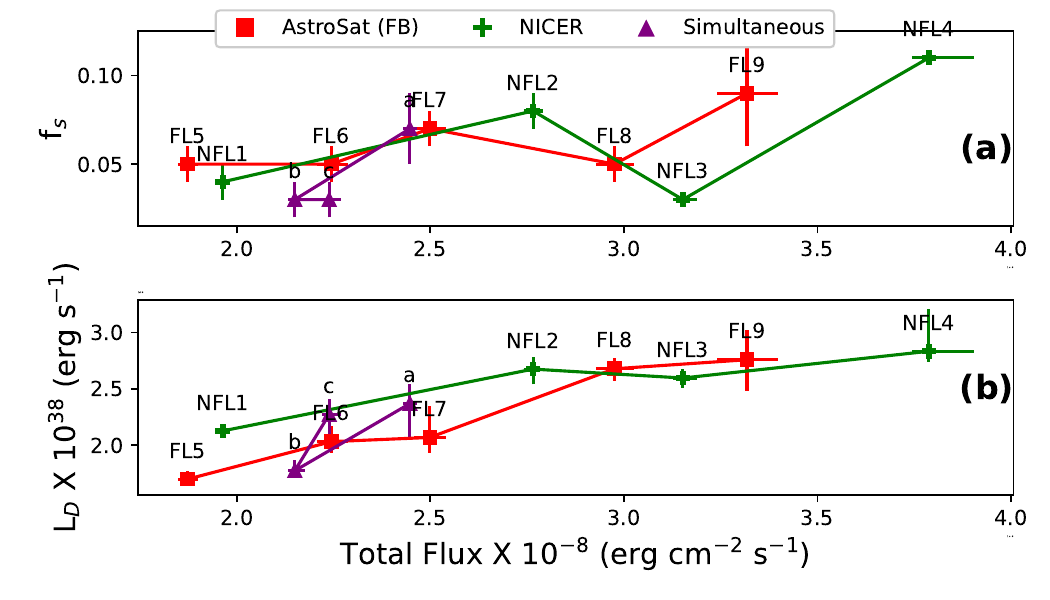}
    \caption{ Variation of spectral features with total unabsorbed flux using the model combination {\tt tbabs*(bbodyrad+thcomp*diskbb)}, where `+' data points represent parameter estimates from NICER observations, while square and triangular data points correspond to values derived from AstroSat analysis of FL 5 to 9 and the AstroSat-NICER joint spectral fitting of simultaneous observations, respectively. The plot shows consistent representative source characteristics, (a) f$_s$ and, (b) L$_{\rm D}$ in erg s$^{-1}$, in all three curves. }
    \label{fig:asni_vartf}
\end{figure}

We investigate the spectral parametric variation from the AstroSat and NICER flux-resolved analysis with the Flux$_{\rm tot}$, as shown in Figures~\ref{fig:as_vartf} and \ref{fig:asni_vartf}, respectively. It shows how the spectral parameters are varying with the increase of X-ray flux along the Z-track. Figure \ref{fig:asni_vartf} reveals the coordination of the NICER spectral parameter with the AstroSat spectral properties from the segment SA (F5) towards FB.

\begin{figure}
	\centering
	\includegraphics[height=6cm, width=8.2cm]{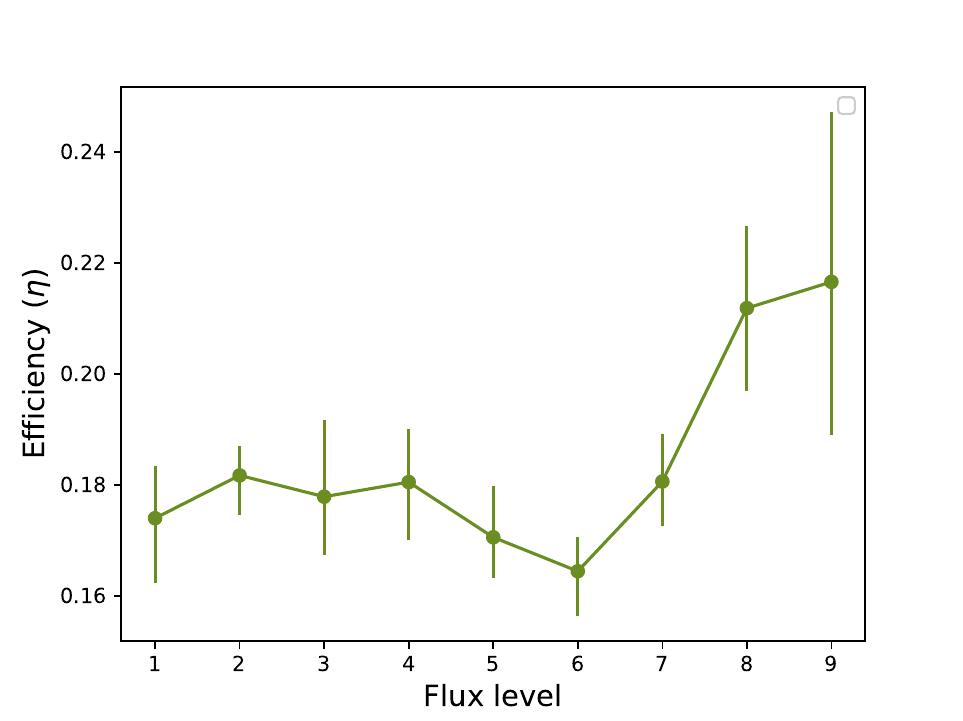}
    \caption{Mass accretion efficiency ($\eta$) of the source using the model combination {\tt tbabs*(bbodyrad+thcomp*diskbb)}, along the flux levels observed with AstroSat }
    \label{fig:eff}
\end{figure}

We estimated the mass accretion efficiency ($\eta$) along with uncertainty of the source along its track. The efficiency, $\eta$ represents the fraction of the accreted mass falling onto the NS surface, which is converted into radiative emission. It is calculated using the relation, \begin{equation}
\eta = \frac{L_{\text{T}}}{\dot{M} c^2}
\end{equation} where L$_{\rm T}$ is the total luminosity of the source, defined as L$_{\rm T}$ = Flux$_{\rm tot}$4$\pi$D$^2$ with D as the distance to the source; and c is the speed of light. The estimated L$_{\rm T}$ for all the FLs of AstroSat are shown in Table~\ref{tab:tab3}.  Figure~\ref{fig:eff} portrays that despite the large variation in $\dot{M}$ and R$_{\rm in}$ seen for the different flux states, the $\eta$ turns out to be nearly constant at $\sim 0.18$ as expected from NS system. This provides support to the spectral model used and validates that the $\dot{M}$ inferred is indeed the correct physical one.

\subsection{Spectral evolution along the flux levels }\label{sec:spvar}
\begin{figure*}
\center
\includegraphics[scale=0.52, angle=0]{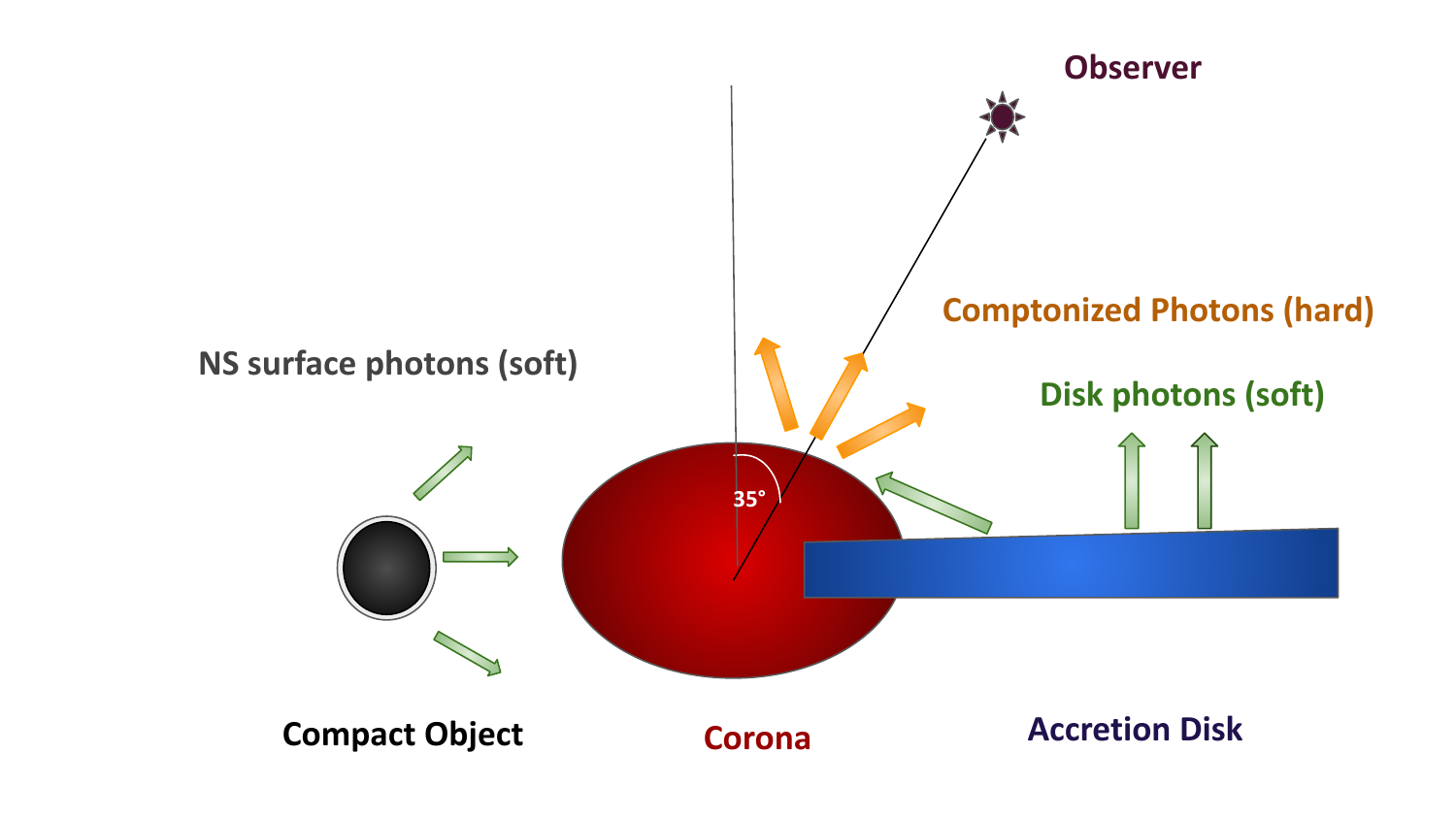}
\caption{\label{fig:geo} {The schematic digaram of the geometry of the system. }}
\end{figure*}

As discussed in Section~\ref{sec:sa}, the source spectrum is well described with thermal component {\tt diskbb}, {\tt bbodyrad}, and one non-thermal component {\tt ThComp}. This suggests seed photon emission from both the NS surface and the accretion disk. Figures~\ref{fig:as_vartf} and \ref{fig:asni_vartf} and Tables~\ref{tab:tab3} and \ref{tab:tab4} represent the spectral parameter variation along the FLs using AstroSat and NICER flux-resolved spectroscopy, respectively, using the primary model combination. Figure~\ref{fig:as_vartf}(a) shows the fractional scattering (f$_s$) is sharply decreasing by $\sim$50\% along the NB, following a similar pattern as observed in the AstroSat HID (see Figure~\ref{fig:obshid}(a)). We observe a low fraction of Comptonized seed photons and Figure~\ref{fig:geo} shows probable geometry of the system, where we can see that some part of the photons from the disk may escape without being significantly Comptonized due to nearly face-on view, and hence giving us a lower fraction of the photons that are being Comptonized. The accretion disk appears to be moving closer towards the central compact object, primarily along the NB as indicated by the observation that the inner-disk temperature (T$_{\rm in}$) is increasing quickly along NB (see Figure~\ref{fig:as_vartf}(c)) while R$_{\rm in}$ decreases correspondingly (see Figure~\ref{fig:as_vartf}(d)) since, R$^{-3/4}$$\propto$ T. As seen in the Figure~\ref{fig:as_vartf}(f), we also noted L$_{\rm D}$, to rise in each FL from FL 1 to FL 9. The $\dot{M}$ is observed vary along the track, with a notable drop at the FL 5 or the SA region (see Figure~\ref{fig:as_vartf}(e)), indicating that $\dot{M}$ might play an important role in traversing the source along the FB.
Figure~\ref{fig:asni_vartf} shows that the spectral parameter variation using NICER is endorsing the AstroSat spectral analysis result, showing identical behavior between NFL 1 to NFL 4 of NICER and FL 5 to FL 9 of AstroSat. Due to decent sensitivity of the NICER, the parameters are well constrained as framed in Table~\ref{tab:tab4} and Figure~\ref{fig:asni_vartf}. Furthermore, we compare the estimated parameters, $\dot{M}$ and R$_{\rm in}$ with the previous studies. Such as, using the Suzaku observations of GX 17+2, \citet{2010Cackett} applied relativistic {\tt diskline} model and estimated the R$_{\rm in}$ to be $\sim$ 7-8 GM/c². NuSTAR spectra of GX 17+2 in the 3-30 keV region were analyzed by \citet{2017Ludlam}, concluding that the disk extended to 1.0-1.02 ISCO. \citet{2019Sriram} estimated the inner disk radius of the source along the track to be within 20–35 km using RXTE data. Based on AstroSat/LAXPC data, \citet{2020Agrawal} estimated R$_{\rm in}$ $\sim$ 28-42 km, which is in well agreement with our estimated range of $\sim$ 30-50 km along the track in the AstroSat analysis. While using LAXPC/SXT spectra \citet{2020Malu} estimated the $\dot{M}$ in NB within the range $\sim$ 2.36-3.42 $\times$ 10$^{18}$ gm/s using the boundary layer luminosity of the source. The value of $\dot{M}$ we estimated using AstroSat observation in NB is $\sim$2.70--2.97 $\times$ 10$^{18}$ gm/s, which is consistent with the reported value.

\subsection{ AstroSat-NICER simultaneous observations}\label{sec:simul}
As indicated in  Table~\ref{tab:1}, AstroSat and NICER has three sets of observations with same dates of observation: September 10, 2018 (hereafter, a), August 14 (hereafter, b), and September 8 (hereafter, c), 2020. Simultaneous observations are widely used to endorse the spectral parameter comparison with different misisons; a few recent studies are \citet{2021Ludlam, 2021Jithesh, 2022Bhargava, 2023Moutard, 2024aBhattacherjee, 2024Combi}. The near simultaneous part of the observation (a) lies mostly in FL 2 and FL 3 in AstroSat and NFL 2 in NICER; for observation (b) it is FL 5 for AstroSat and NFL 4 in NICER; for observation (c) it is in FL 5 for AstroSat and NFL 1 for NICER. We conducted joint spectral fitting between NICER/XTI and AstroSat/LAXPC-SXT spanning the energy range of 0.5–20 keV utilizing the synchronous data, to ensure the consistency of the results with the spectral analysis along FLs. Figure~\ref{fig:asni_vartf} shows the consistency between the joint fitting of the two missions and the flux-resolved spectroscopy of NICER and AstroSat spectral parameters. 

\subsection{ Alternate spectral model combination: Comptonization of blackbody emission }\label{sec:alt}
\begin{figure}
	\centering
	\includegraphics[height=20cm, width=8.5cm]{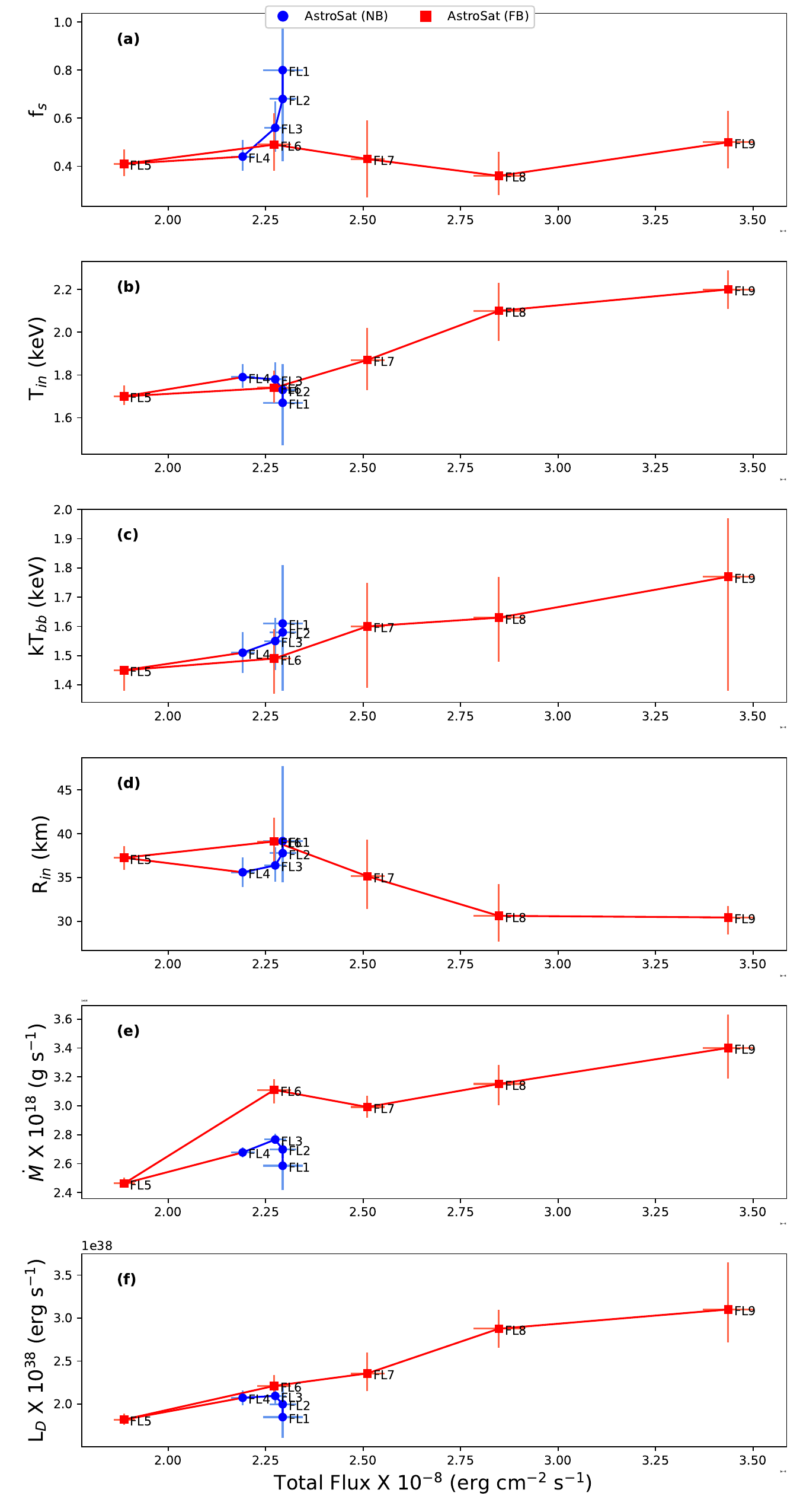}
    \caption{Variation in the source's spectral characteristics, obtained from flux-resolved spectroscopy using the model combination {\tt tbabs*(diskbb+thcomp*bbodyrad)}, relative to the total unabsorbed flux across different FL. The circular and square data points represent the source's properties across FL 1 to 4 (NB) and FL 5 to 9 (FB), respectively as observed by AstroSat. The plot reveals distinct change in behavior of the source from NB to the FB. The plot parameters are organized as follows: (a) covering fraction, f$_s$; (b) inner-disk temperature, T$_{\rm in}$ in keV; (c) blackbody temperature, kT$_{bb}$; (d) inner-disk radius, R$_{\rm in}$ in km; (e) mass-accretion rate, $\dot{M}$ in gm/s; (f) disk luminosity, L $_{\rm D}$ in erg s$^{-1}$, respectively. }
    \label{fig:as_vartf2}
\end{figure}

\begin{figure}
	\centering
	\includegraphics[height=6cm, width=8.2cm]{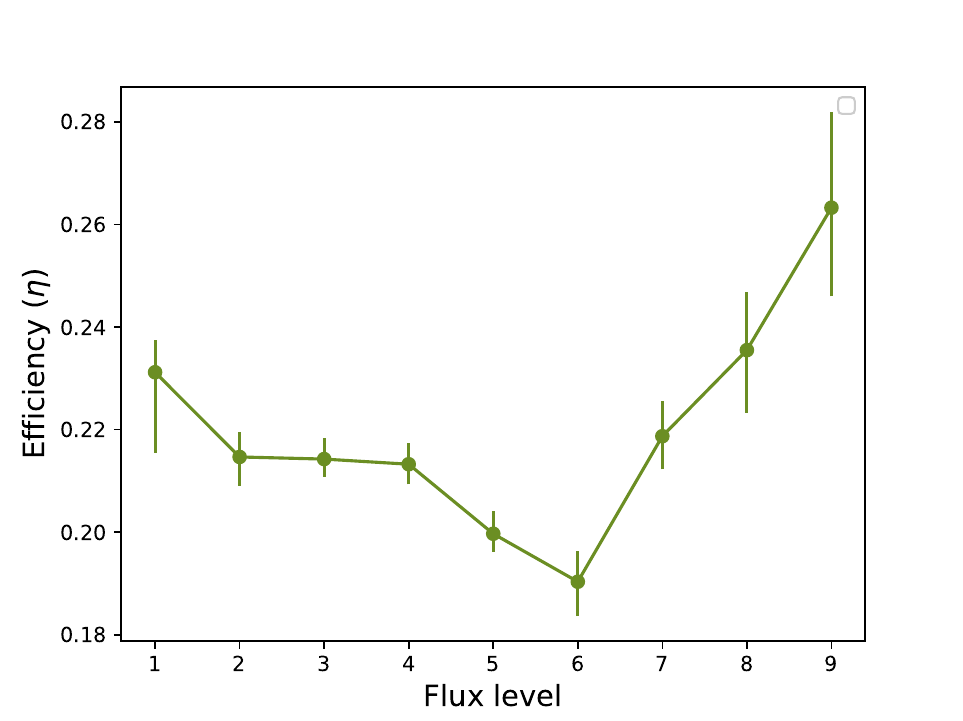}
    \caption{Mass accretion efficiency ($\eta$) of the source using the model combination {\tt tbabs*(diskbb+thcomp*bbodyrad)}  along the flux levels observed with AstroSat.}
    \label{fig:eff2}
\end{figure}

As discussed in Section~\ref{sec:aa}, we tested the alternative spectral model combination {\tt tbabs*(diskbb+thcomp*bbodyrad)}, which resulted in a similar $\chi^2_{\rm red}$ value (see Figure~\ref{fig:rchi}), with N$_{\rm bb}$ fixed corresponding to 10 km. Figure~\ref{fig:as_vartf2} illustrates the relative variation of the key parameters using the alternate model. The trends in the source characteristics such as f$_{\rm s}$, T$_{\rm in}$, $\dot{M}$ and L$_{\rm D}$ along the FLs are closely comparable with those observed in the primary model, {\tt tbabs*(bbodyrad+thcomp*diskbb)}. For example, in NB, the $f_{\rm s}$ plays a crucial role in driving the source along the branch, decreasing by $\sim$ 50\%, while it becomes nearly constant in the FB. The other source parameters like $\dot{M}$ (which increases by $\sim$ 40\%, from SA to FB), L$_{\rm D}$, T$_{\rm in}$ exhibit similar trend along the track. Figure~\ref{fig:eff2} illustrates that in this model, the accretion efficiency $\eta$ also remains consistent $\sim$ 0.22, a physically justified value for a binary system with a NS as the accretor. However, the behavior of R$_{\rm in}$, differs in those two models. In the primary model, R$_{\rm in}$ undergoes substantial changes along the NB, making it one of the governing factor in tracing the source's spectral evolution.  In contrast, when NS surface photons are assumed to be Comptonized, R$_{\rm in}$ remains nearly constant ($\sim$30–40 km) throughout the track.  This range is consistent with previously reported ranges and aligns with our primary model well, where Comptonization is caused by accretion disk photons. 
While the inferred physical properties are in agreement within the model framework, a more detailed investigation-- incorporating data over a wider energy range and including multiwavelength observations—could provide additional constraints, further improving the understanding of the underlying physical mechanisms. 

\section{Discussion} \label{sec:d}

In this study we carried out a broad-band spectral analysis of the source GX 17+2, using all publicly available observations of AstroSat/LAXPC-SXT and NICER/XTI between the year 2016 to 2020. A wide energy coverage aids in the detailed characterization of components, including the accretion disk, corona, and compact object. Furthermore, it allows for better constraints on the physical parameters.

To understand the formation of Z-shaped track has mostly been the focus of the comprehensive studies of Z-sources (example, GX 349+2: \citet{2018Coughenour, 2023Kashyap}; GX 17+2: \citet{2002Homan, 2012Lin, 2020Agrawal}; GX 340+0: \citet{2006Church, 2013Seifina, 2023Bhargava, 2024Chattopadhyay}; Cyg X-2: \citet{2002DiSalvo, 2006Church, 2010Church}; Sco X-1:\citet{2003Bradshaw, 2007Dai}; GX 5-1: \citet{2002Jonker, 2019Bhulla, 2024P}). As discussed in Section~\ref{sec:intro}, the Sco-like sources traces a prominent NB and FB while in case of Cyg-like sources it traces HB and NB with a faint FB. After rigorously studying these sources, \citet{2012Church} proposed that the cause of formation of the FB in Cyg-like sources are due to the unstable nuclear burning; while in Sco-like sources it is associated with the combination of unstable nuclear burning and increase of $\dot{M}$ resulting strong flares. From our analysis we find similar behavior as shown in Figure~\ref{fig:as_vartf}(e) revealing a sudden increase of $\dot{M}$  by nearly $\sim$ 40\% after a drop in NB-FB vertex, signifying a change in the physical process associated with flaring state of the source, showing a correlation along the Z-track. 
However, \citet{2012Lin} on performing detail S$_z$ resolved spectroscopy of GX 17+2 using RXTE data showed that $\dot{M}$ is constant throughout the Z-track. The authors obtained the variation of R$_{\rm in}$ to sharply decline along FB while is nearly constant in NB. However, in our analysis, we observe a declining behavior of R$_{\rm in}$ in NB while without changing significantly in FB, making it one of the factor responsible for the source to traverse in NB (refer Figure~\ref{fig:as_vartf}(d)). Another factor responsible for the source to trail in NB is the fraction of disk photons entering the corona for the Comptonization or the f$_{\rm s}$, which significantly decrease by $\sim$50\% along NB and becomes nearly constant in FB (refer Figure~\ref{fig:as_vartf}), signifying its prominent role in NB. Even though there is distinct variation of the parameters in different spectral states, $\eta$ is nearly constant for all the HID segments (FL 1 to FL 9) at $\sim$0.20, which is expected for a NS system. Hence, it can also be interpreted that the estimated distance of the source which is determined to be 13 kpc, without any optical counterpart, by \citet{2008Galloway} based on observations of the thermonuclear burst of the source—is plausible and is consistent with the analysis.

This work is primarily focused on the flux co-relation study of the Sco-like Z-source, GX 17+2. Earlier, similar flux co-relation study using AstroSat/LAXPC-SXT observations has been carried out for Cyg-like source: GX 5-1 \citep{2019Bhulla} and GX 340+0 \citep{2024Chattopadhyay}.

\citet{2019Bhulla} segmented the HID of the source GX 5-1 into segments and performed the spectral analysis using strict simultaneous data of LAXPC and SXT, defining the X-ray spectrum with multicolor blackbody emission and a Comptonized emission component. From the analysis, the authors obtained disk flux ratio to be the only parameter to exhibit co-related variation with the different segments, inferring it to act as the driver to regulate different position of the source in the Z-path. 
However, \citet{2024Chattopadhyay}, explained their X-ray spectrum of GX 340+0 with a single temperature blackbody component and a Comptonized component. In each HID segments, they perform the spectral analysis independently, the spectral parametric variation are then compared with the blackbody flux ratio (blackbody flux upon Flux$_{\rm tot}$), w.r.t. to which, a decrease in blackbody radius and an increase in blackbody temperature are observed along the segments, while covering fraction shows a steady decline. The authors noticed a complex behavior of blackbody flux and Comptonization flux to understand the factor responsible for the Z-track formation. Notably, \citet{2023Bhargava} also explicitly studied the spectral evolution of the source GX 340+0 using AstroSat observations and investigated the spectral variation along the track from HB to NB. Their spectral decomposition study also suggests that the interplay of both blackbody and Comptonizing components help evolution of HB and NB. 

\citet{2023Kashyap} investigated the spectral evolution of the Sco-like source GX 349+2 using data from AstroSat and NICER, focusing on the transition from the SA to the FB. They analyzed the spectral variation relative to the total LAXPC flux (4–25 keV) and NICER flux (0.7–8 keV) along the track. The X-ray spectrum was found to be well-represented by a combination of blackbody radiation, a multicolor blackbody model, and a non-thermal power-law component. A relatively low flux was detected in the SA compared to the FB. Additionally, a decrease in the R$_{\rm in}$ along the NB/FB to FB, accompanied by an increase in blackbody temperature, indicates that the source is moving closer to the compact NS \citep{2002Frank}, similar to the behavior we obtained for T$_{\rm in}$ and R$_{\rm in}$ (see Figure~\ref{fig:as_vartf}(c \& d)). \citet{2023Kashyap} compared their result from AstroSat observations with NICER observation reporting more constrained parameter values. Likewise we observe similar trends with AstroSat observation and constrained values of the parameters on comparing with the NICER observation as shown in Figure~\ref{fig:asni_vartf}. We also used the AstroSat-NICER simultaneous data to confirm the consistency of the results (see, Figure~\ref{fig:asni_vartf}). 

In this work, we discussed the spectral result from two combinations of simple model, the {\tt tbabs*(bbodyrad+thcomp*diskbb)} and {\tt tbabs*(diskbb+thcomp*bbodyrad)}. In both the cases, we obtained comparable relative variation of the parameters as well as the source characteristics. For instance, in NB, the $f_{\rm s}$ plays the key role for the source to trail in the branch as it decrease by $\sim$ 50\%, while it becomes nearly constant in the FB. The other source parameters like $\dot{M}$ (which increases by $\sim$ 40\%, from SA to FB), L$_{\rm D}$, T$_{\rm in}$ shows similar trend along the track. Unlike R$_{\rm in}$, which is obtained to be significantly decreasing in NB and then becoming constant in FB in the primary model. While in the alternate model it prevails within the range of $\sim$ 30--40 km throughout the track, though in both the cases its values remain within the previously reported range for the source. 

Hence, this work highlights the requirement of broadband spectral analysis to probe the spectral behavior more precisely and widely. It also marks that the parameters undergo distinct variation from the SA region following the extended  FB track, suggesting its association with the change in the disk structure and the accretion process as well. Hence, to understand the formation of the Z-tracks more accurately, extensive study of the vertex regions i.e. HA (HB/NB) and SA (NB/FB) of the Z-sources is essential. Furthermore, studying such sources over a broader energy range, incorporating aspects like polarization and multi-wavelength observations, would provide more insights into the underlying physical mechanisms. 
\section{Conclusion} \label{sec:conc}

In this work, using all the AstroSat and NICER observations from 2016 to 2020, we conducted a comprehensive broadband spectral correlation study of the Sco-like Z-source GX 17+2. Our study reports the variation in the source's spectral properties across different intensity levels, facilitating an understanding of the source's behavior in relation to X-ray flux and, consequently, the physical processes directly associated with accretion. The study reveals that properties such as the covering fraction, with a substantial decrease of $\sim$ 50\% and R$_{\rm in}$, apparently approaching closer to the central compact object in NB, leading to an increase in T$_{\rm in}$, suggests a correlation with the source’s movement along the NB. We also observe a substantial variation of the total luminosity, L$_{\rm T}$ from $\sim$ 4.0 to $\sim$7.0 $\times 10^{38}$ ergs s$^{-1}$ along the branch. Despite the significant variation correlated with L$_{\rm T}$, $\dot{M}$, and R$_{\rm in}$, we obtained the accretion efficiency constant at $\sim$ 0.20 as supposed for a NS system, supporting the spectral model we applied. This analysis also supports the distance of the source to be $\sim$13 kpc as determined by \citet{2008Galloway}, without any optical counterpart, since the $\eta$ is proportionally correlated with the source distance.  The NICER data and the AstroSat-NICER near simultaneous data provides a more constrained spectral properties, explaining and satisfying the variation clearly. Hence this study underscores the need for broadband studies of the vertex regions and the branches as well, to address the decades-old goal of understanding the formation of the Z-track, and the associated accretion processes.

\section*{ACKNOWLEDGMENTS}

This work has utilised the softwares provided by the High Energy Astrophysics Science Archive Research Centre (HEASARC). This research utilized the LAXPC and SXT instruments data of AstroSat mission of ISRO. We gratefully acknowledge the LAXPC and SXT Payload Operation Centers (POC) at TIFR, Mumbai, for providing the data via the ISSDC data archive and for providing the necessary software tools. We also thank XTI/NICER team, for providing the analysis guidelines in their website. S.B. acknowledges the periodic visits to Inter-University Centre for Astronomy and Astrophysics (IUCAA), Pune, to carry out substanctial portion of this work. B.S. acknowledges the IUCAA Visiting Associateship Program and thanks IUCAA for their hospitality during his visits, where part of this research was conducted.  This research has made use of MAXI data provided by RIKEN, JAXA and the MAXI team \citep{2009Matsuoka}. S.B. acknowledges the financial support provided by the INSPIRE fellowship (Grant No.: DST/INSPIRE Fellowship/[IF220164]) from the Department of Science and Technology (DST), Ministry of Science and Technology, India. V.J. acknowledges the support provided by the Department of Science and Technology (DST) under the `Fund for Improvement of S \& T Infrastructure (FIST)' program (SR/FST/PS-I/2022/208). V.J. also thanks IUCAA, Pune, India, for the Visiting Associateship. Finally, we thank the anonymous referee for providing useful comments and suggestions, which considerably enhanced the paper’s content.


\end{document}